\documentclass{emulateapj}

\newcommand{\G}{\Gamma}

\newcommand{\sT}{\sigma_{\rm T}}

\newcommand{\e}{\epsilon}
\newcommand{\g}{\gamma}
\newcommand{\gp}{\gamma}

\newcommand{\Vp}{V^\prime}

\newcommand{\xD}{x_\Delta}
\newcommand{\Op}{\Omega^\prime}

\newcommand{\tp}{t^\prime}
\newcommand{\ep}{\epsilon^\prime}
\newcommand{\Dop}{\delta_{\rm D}}

\newcommand{\ti}{\theta_i}
\newcommand{\tcl}{\theta_{cl}}
\newcommand{\psim}{\lower.5ex\hbox{$\; \buildrel \propto \over\sim \;$}}
\newcommand{\lbar}{\lower.0ex\hbox{$\; \buildrel {\lower0.0ex \hbox{-}} \over\lambda  \;$}}

\shorttitle{GRBs and the Swift Observations}
\shortauthors{Dermer}

\begin{document}
\title {Nonthermal Synchrotron Radiation from Gamma Ray Burst
External Shocks and the X-ray Flares Observed with Swift}

\author{Charles D.\ Dermer}
\affil{U.S.\ Naval Research Laboratory, Code 7653, 4555 Overlook SW, 
Washington, DC,
  20375-5352} 
\email{charles.dermer@nrl.navy.mil}

\begin{abstract}

An analysis of the interaction between a spherical relativistic 
blast-wave {\it shell} and a stationary {\it cloud} with a spherical cap
geometry is performed assuming that the cloud width $\Delta_{cl}\ll
x$, where $x$ is the distance of the cloud from the GRB explosion
center.  The interaction is divided into three phases: (1) a collision
phase with both forward and reverse shocks; (2) a penetration phase
when either the reverse shock has crossed the shell while the forward
shock continues to cross the cloud, or vice versa; and (3) an
expansion phase when, both shocks having crossed the cloud and shell,
the shocked fluid expands. Temporally evolving spectral energy
distributions (SEDs) are calculated for the problem of the interaction
of a blast-wave shell with clouds that subtend large and small angles
compared with the Doppler(-cone) angle $\theta_0 = 1/\Gamma_0$, where
$\Gamma_0$ is the coasting Lorentz factor.  The Lorentz factor
evolution of the shell/cloud collision is treated in the adiabatic
limit.  Behavior of the light curves and SEDs on, e.g., $\Gamma_0$,
shell-width parameter $\eta$, where $\Delta_0 +
\eta x/\Gamma_0^2$ is the blast-wave shell width, and properties and locations of the
cloud is examined.  Short timescale variability (STV) in GRB light
curves, including $\sim 100$ keV $\gamma$-ray pulses observed with
BATSE and delayed $\sim 1$ keV X-ray flares found with Swift, can be
explained by emissions from an external shock formed by the GRB blast
wave colliding with small density inhomogeneities in the ``frozen
pulse" approximation $(\eta \rightarrow 0$), and perhaps in the
thin-shell approximation $(\eta \approx 1/\Gamma_0$), but not when
$\eta \approx 1$.  If the frozen-pulse approximation is valid, then external shock
processes could make the dominant prompt and afterglow emissions in
GRB light curves, consistent with short delay two-step collapse models
for GRBs.

\end{abstract}


\section{Introduction}

Important knowledge about the nature of GRBs comes from analyses of
burst light curves. BATSE, triggering on peak flux over 64, 256, and
$1024$ ms timescales in the 50 -- 300 keV band, provides the largest
database at hard X-ray/soft $\gamma$-ray energies, amounting to over
2292 GRBs reported nearly 8 years into the mission \citep{fis99}, and
$\sim 2700$ BATSE GRBs in total.  Examining BATSE GRB light curves
gives the impression that the $\gamma$-ray activity, so intense during
the first $\sim 1$ -- 100 s, begins to weaken with time, all the while
displaying pulses with relatively constant widths. If this evidence
requires for its explanation a very powerful structured relativistic
wind from an active central engine that becomes less vigorous with
time, then the implied GRB explosion mechanism is very different than
if this data is taken as evidence for a single explosive event where
the GRB pulses are made by external shocks from a weakening blast wave
that interacts with material in the circumburst medium.

The GRB afterglow revolution, initiated by Beppo-SAX and continued
with HETE-2 and INTEGRAL, gave for the first time GRB source redshifts
and distances and therefore apparent isotropic powers and
energies. With the opening angle of the relativistic jet inferred from
the achromatic beaming breaks in optical light curves, the absolute
energy of a GRB is fixed to the uncertainty in the efficiency of
radiation production.  The tools of standard relativistic blast-wave
physics for the afterglow external shock emissions allow one to test
for uniform or wind-formed circumburst media, and determine the $\e_e$
and $\e_B$ parameters for assumed jet structures \citep{pk02}.
Fitting the statistical data on the GRB redshift and opening angle
distributions can determine the intrinsic jet opening angle and
whether a top-hat model for GRB jet structure gives good agreement
with the data \citep{gp06,ld07}.

A rich new database is recently opened by Swift \citep{geh04}. After
triggering with BAT in the 15 -- 150 keV range, Swift autonomously
slews within $\approx 100$ s to a burst so that the XRT, with
excellent sensitivity, can get a clear picture of the evolution of the
0.3 -- 10 keV X-ray emission.  In large numbers of GRBs, there are
steep declines in the X-ray emission at the end of the prompt X-ray
luminous phase hundreds of seconds after the start of the GRB
\citep{tag05}, an extended plateau phase \citep{bar05}, and X-ray
flares \citep{bur05}, which in some cases have durations that are a
small fraction of the time since the start of the GRB.  Extrapolation
of the BAT into the XRT band gives an almost continuous X-ray light
since the GRB trigger \citep{obr06}.

The interpretation of this data is complicated not only by the rich
detail revealed by Swift but, for statistical studies, by its unusual
triggering method, which makes the assignment of a threshold flux
problematical \citep{ban06}.  Because the Swift BAT uses a coded mask
from which images of the GRB photons can be formed, a lower peak flux
than pre-Swift telescopes can be identifiable as a GRB by using both
rate and image triggers.  A broader dynamic range, from 0.004 s to 26
s, is used for the rate trigger. Differences in redshift distributions
of Swift GRBs compared with the GRBs detected with BATSE, Beppo-SAX
and HETE-2, which had comparable triggering criteria
can, however, mainly be understood by different peak flux
sensitivities of Swift and pre-Swift detectors \citep{ld07}.
 
The X-ray flares in the late prompt and early afterglow $(t\sim 10^2$
s -- $10^4$ s) observed with the Swift satellite have been interpreted
as evidence for internal shocks and refreshed energization of the
relativistic outflow \citep[][]{zha06,chi07}. \citet{lia06} and
\citet{yam06} model the steeply declining phase as a superposition of
background external shock emission and a curvature pulse produced by
shells ejected long after the start of the GRB. This model for energy
dissipation in GRBs and production of the GRB light curves is
commensurate with the collapsar/failed supernova scenario for
long-duration GRBs,\footnote{Long-duration GRBs are referred to
henceforth as GRBs; here we do not consider the short hard GRB class
or the low-luminosity GRBs, also argued to form a separate class
\citep{lzv06}.}  where the evolved iron core of a massive star
collapses to a black hole. In the collapsar picture, energy
dissipation in an accretion torus or through the Blandford-Znajek
process drives a narrow collimated jet through the stellar
envelope. During the early prompt phase, continued activity drives
successive waves of collimated relativistic plasma through the SN
shell. Collisions between the shells make GRB pulses. Later activity
and peculiarities in the light curves are attributed to refreshed
outflows or internal shocks from late outbursts of the
engine. Interpreting rapid variations in the GRB X-ray light curve as
activity of the central engine, then the engines in some GRB sources
must be on for rest-frame times $\lesssim 10^4$ s after the start of
the event \citep{bur05,fal06,bur05a}.

Early calculations of the spectral and temporal behavior of the GRB
pulses in an internal shock model were made by \citet{dm98},
\citet{kps97}, and \citet{gps99}. The efficiency difficulties for the
internal shock model \citep{kum99,bel00,mam07} are reduced by large
Lorentz-factor contrasts between different shells \citep{kp01}, which
constrains the collision radius and $\nu F_\nu$ peak energies under
the requirement of Thomson thinness \citep{gsw01}.  For the internal
shock/colliding shell model, as well as for the external shock model,
efficiency concerns are ameliorated by the large hadronic energy
content in the prompt phase that escapes to form the ultra-high energy
cosmic rays \citep{der06}.

A complete description of the collision between two relativistic
shells that follows both the hydrodynamics and radiation physics
needed to calculate emission spectra and light curves from different
pulses in GRBs, necessary to test the internal shell model, has
only been recently treated, 
in a series of papers by \citet{mim04, mam05,mam07}.  
Here we present a
complete and systematic analysis of the collision between a cold GRB
blast wave shell and a cloud with a spherical cap geometry in an
external shock model.  We use these results to check the widespread claims
by the Swift team that central engine activity is responsible for the
erratic X-ray flares from GRBs
\citep[e.g.,][]{obr06a,fal06a,rom06,bur07}.  We dispute these claims
and argue that an external shock model can make the prompt
$\gamma$-ray pulses and afterglow X-ray flares under the condition of
no spreading of the cold blast-wave fluid shell.  If this assumption
is allowed, then interesting implications for the nature of the
collapse process in GRB sources follow. A short delay two-step
collapse process, in fact, provides a competing and compelling
alternative to the collapsar/failed supernova model, as discussed
below (Section 5).

The emissivities calculated from the interaction of a blast-wave shell
and a stationary cloud are integrated over three spatial dimensions to
obtain spectral fluxes as a function of observer time. Only the
synchrotron component is considered here \citep[for a recent treatment of 
the synchrotron-self Compton component in external shocks, see][]{gp07}, 
and other assumptions are
made to simplify the analysis, for example, a randomly ordered
magnetic field and isotropic electron distribution in the proper frame
of the shocked fluid.

The techniques of the standard blast-wave model are used
\citep[e.g.][]{spn98,pir99,pir05,mes06}, and the results are
applied to observations of prompt $\sim 100$ keV emission pulses and
$\sim 1$ keV X-ray flares in GRB light curves. The problem is analyzed
and broken into three distinct interaction phases, as described in
Section 2.  Calculations of light curves and SEDs from the emissions
of external shocks formed by the interaction of a GRB blast wave with
a stationary cloud are presented in Section 3. Parameter are chosen to
produce short timescale variability (STV; $\Delta t_v/\hat t \ll 1$,
where $\Delta t_v$ is the variability time scale and $\hat t$ is the
time since the start of the GRB) in the $\gamma$-ray lightcurve, which
is only possible if the blast wave shell width can be approximated by
a a thin shell ($\Delta(x) \lesssim x/\Gamma^3_0$) or frozen
(unspreading) shell.  This result is demonstrated analytically in
Section 4, and illustrated by Monte Carlo simulations of GRB light
curves.  We conclude that the X-ray flares observed with Swift are
compatible with an external shock model if the GRB blast wave shell
interacts with a clumpy surrounding external medium and does not
spread transversely. This assumption is discussed in Section 5, and is
argued to be correct. Implications for the nature of the collapse
mechanism and jet formation in long-duration GRBs are also discussed
in Section 5.  The study is summarized in Section 6.

The reader who is not interested in the detailed analysis may avoid
Section 2, and refer directly to the numerical results in 
Section 3. The principal conclusions of the analysis are found in the
simplified analytic descriptions in Section 4.

\section{Analysis of the Blast-Wave/Cloud Interaction}

Consider a GRB that takes place at redshift $z$ and releases energy
over a characteristic timescale $\Delta_0/c$ representing the period
of activity of the GRB central engine.  The corresponding spatial
dimension $\Delta_0$ should probably be greater than the
Schwarzschild radius of the several Solar mass black hole formed in
the GRB event; thus $\Delta_0\gtrsim 10^6~{\rm cm}$.  Depending on the
nature of the central engine of GRBs, $\Delta_0$ could range from a
fraction of a second to hundreds of seconds or longer if, as in the
collapsar model for the GRB prompt phase, the duration of the highly
variable X-ray and $\gamma$-ray flux in a GRB is assumed to reflect
the period of activity of the engine.  Here we consider a GRB
engine where the progenitor neutron star collapses impulsively to a
black hole, so that $\Delta_0 \sim 10^{7}\Delta_{7}$ cm, with
$\Delta_{7}\sim 1$.

The apparent isotropic equivalent $\gamma$-ray energy released by a
GRB explosion is written as $E_0$.  The apparent
20 keV -- 2 MeV isotropic rest-frame energy measured from GRBs
typically range in values from $\approx 10^{51}$ -- $10^{54}$ ergs,
with a handful of anomalously low energy GRBs with energies as low as
$10^{48}$ ergs \citep[e.g., GRB 980425;][]{fb05,gngf07}.  A
significant number of GRBs have $E_0\gtrsim 10^{54}$ ergs, so that the
total GRB energy must be larger, by a factor of at least several (and
possibly much larger if the radiation efficiency is low or energetic
hadrons are formed and escape, e.g., as neutrons \citep{ad03,der06}).
In this study, jet effects are considered as a restriction on the
interaction angle, the jet structure is assumed to be `top-hat,' and
there is no lateral and very little transverse spreading.

The explosion is assumed to form a fireball with initial Lorentz factor 
(entropy per baryon) denoted by
\begin{equation} 
\Gamma_0 = \;{E_0\over M_0 c^2}\; \gg 1\;,
\label{M_0} 
\end{equation} 
where $M_0$ is the amount of baryonic matter mixed into the initial
explosion. In this analysis, we do not consider effects
of neutron decoupling \citep{dkk99,bm00,bel03}, and furthermore assume
that the blast-wave energy in the coasting stage is carried primarily
by particle rather than field energy \citep[see, e.g.,][for the latter
case]{lb01}.

\begin{figure}[t]
\epsscale{1.0}
\plotone{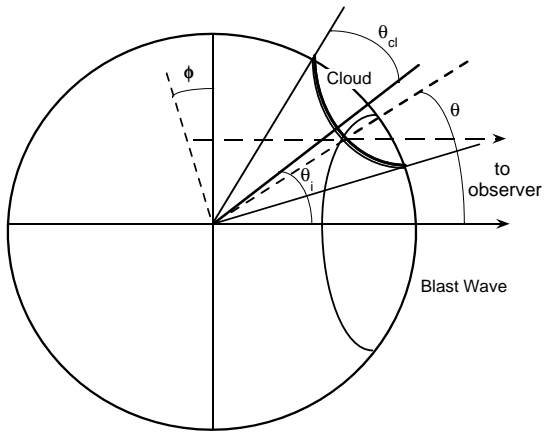}
\caption{ Illustration (not drawn to scale) of the geometry of a 
spherical, relativistic blast-wave shell interacting with a cloud
having a spherical cap geometry. The axis through the center of the
cloud is offset from the observer's line of sight by the angle
$\theta_i$.  The cloud subtends an angle $\theta_{cl}$ as measured
from the explosion center. Emission originates from radiating plasma
located at a distance $x$ from the center of the GRB explosion, and at
angles $\theta$ and $\phi$ measured with respect to the line-of-sight
direction to the observer. To approximate a quasi-spherical cloud, the
transverse and lateral extents of the cloud are set equal.  }
\label{f1}
\end{figure}

The blast-wave plasma shell reaches its coasting Lorentz factor
$\Gamma_0$ at distance $x \gtrsim \Gamma_0 \Delta_0$ from the
explosion.  At $x\gtrsim x_{spr} \equiv \Gamma_0^2 \Delta_0$, where
$x_{spr}$ is the spreading radius, internal motions within the
blast-wave shell\footnote{The relativistic blast-wave shell from the
GRB will always be referred to as the {\it shell}, and the stationary
material that is swept up by the shell will be referred to as a {\it
cloud}, even when the cloud represents supernova remnant (shell)
material.}  are thought to cause it to spread so that $\Delta(x)
\approx \eta x/\Gamma_0^2$ \citep{mlr93,pir99}, 
where $\eta \lesssim 1$. We consider the case that the shell
experiences little spreading, so that $\eta \ll 1$, which has as its
asymptotic limit the frozen-pulse approximation ($\eta \rightarrow 0$).  
More discussion about $\eta$ and shell spreading is given in Section 4.4.

In the stationary frame of the explosion, the shell width $\Delta(x)$
is therefore given by
\begin{equation} \Delta(x) \cong
 \Delta_0 + \eta\;{x\over \Gamma_0^2}\;.
\label{Delta(x)}
\end{equation}
The proper number density of the relativistic shell is given by
\begin{equation}
n(x) = {E_0 \over 4\pi x^2 \Gamma_0^2 m_p c^2 \Delta(x)} \;.
\label{n(x)}
\end{equation}

\subsection{Geometry of the Blast Wave/Cloud Interaction}

The interaction event is sketched in Fig.\ 1.  The blast-wave shell of
width $\Delta = \Delta(x_0)$ collides with a uniform cloud with
density $n_{cl}$. The cloud is assumed to have a spherical cap
geometry and to subtend a solid angle $\Delta \Omega_{cl} = \pi
\theta_{cl}^2$ as measured from the explosion center, and it therefore
presents a projected area $A_{cl} = \pi \theta_{cl}^2 x_0^2$ to the
GRB blast wave, where $x_0$ is the distance from the origin to the
inner edge of the cloud. The angle between the axis through the cloud
center and the line-of-sight to the observer is denoted $\theta_i$.
The coordinates ${\vec x} =(x,\theta,\phi )$ defining the location of
the radiating material in the stationary (explosion) frame are
measured with respect to the line-of-sight to the observer, with the
angle $\phi$ representing the projection of the azimuth on the plane
normal to the direction to the observer; thus $\phi = 0$ is defined
with respect to the projection of the axis through the cloud center on
this plane (see Fig.\ 1a).  For simplicity, both the shell and cloud
are assumed to be composed of electron-proton plasma.

The $\nu F_\nu$ flux, denoted by $f_\e(t)$, measured at observer time
$t$ and at dimensionless photon energy $\e = h\nu/m_ec^2$, is given by
\begin{equation}
f_\e (t)  =  d_L^{-2}\;\int_0^{2\pi} 
d\phi \int_{-1}^1 d\mu  \int_0^\infty dx\; 
x^2 \ep j^\prime(\e^\prime,\vec\Op; {\vec x},t^\prime)\;\Dop^3\;,
\label{fet1}
\end{equation}
where primes refer to comoving quantities, $\ep= \e_z/\Dop \equiv
(1+z)\e/\Dop$, and the integration is over volume in the stationary
(explosion) frame
\citep{gps99,kp00,der04}. The Doppler factor of the radiating fluid is
defined by the expression
\begin{equation}
\Dop\ ={1\over \Gamma(1-\beta\mu)}\;,
\label{delta}
\end{equation}
where $\mu = \cos\theta$, $\Gamma = \Gamma(\vec x,t_*)$ is the
emitting fluid's Lorentz factor, and $\beta c = c
\sqrt{1-\Gamma^{-2}}$ is its speed. The differential emissivity
$$j^\prime(\e^\prime,\vec\Op ; {\vec x},t^\prime) = {d{\cal E}^\prime
(\vec\Op ) \over d\Vp d\tp d\Op d\ep}$$ is defined such that
$j^\prime(\e^\prime,\vec\Op ; {\vec x},t^\prime)d\Vp d\tp d\Op d\ep$
is the differential comoving energy $d{\cal E^\prime}$ of photons with
dimensionless energies between $\ep$ and $\ep + d\ep$ that are
radiated during time $d\tp$ from comoving volume $d\Vp$ within the
solid angle element $d\Op$ in the direction $\vec\Op$ defined in the
comoving fluid frame. Although the directional properties of the
radiation must be considered in the general case of nonthermal
synchrotron emission from relativistic electrons in an ordered
magnetic field, or for external Compton processes, here we assume that
the radiation is emitted isotropically in the comoving frame, and
assume a nonthermal synchrotron origin for the keV -- MeV emission
from GRBs. Implicit in this approximation is that the magnetic field is
randomly ordered and that the electrons have an isotropic pitch angle
distribution in the fluid frame.

Thus $j^\prime(\e^\prime,\vec\Op ; {\vec x},t^\prime) =
j^\prime(\e^\prime ; {\vec x},t^\prime)/4\pi$. Because the azimuthal
dependence no longer appears in either $j^\prime(\e^\prime ; {\vec
x},t^\prime)$ or $\Dop$, it is trivial to integrate eq.\ (\ref{fet1})
over $\phi$ under the assumption that the emissivity
$j^\prime(\e^\prime ; {\vec x},t^\prime)= j^\prime(\e^\prime ;
x,t^\prime)$, that is, the emissivity depends only on $x$ and $t$ but
not on angle (except as determined by the physical extent of
cloud). This assumption allows us to treat the hydrodynamics of the
blast-wave/cloud interaction in a planar geometry (Section 2.2). We
obtain
$$f_\e (t) = (2\pi d_L^{2})^{-1}\;\int_{\ti - \tcl }^{\ti + \tcl }
d\theta\; |\sin\theta |\;\phi_+
$$
\begin{equation}
\times\int_0^\infty dx\; x^2 \ep j^\prime(\e^\prime;  x,t^\prime)\;\Dop^3\;,
 \label{fet2}
\end{equation}
noting that the Doppler factor is, in general, dependent on
location. The term $\phi_+ = \phi_+(\theta,\tcl,\ti)$ represents the
maximal azimuthal angle subtended by the cloud for received emission
from radiating plasma located at an angle $\theta$ with respect to the
observer's line of sight; see Fig.\ 1.  From the addition formula for
angles in spherical geometry, we have that $\cos\theta_{cl} =
\cos\theta \cos\theta_i + \sin \theta \sin\theta_i\cos\phi_+.$
Inspection of Fig.\ 1 reveals that when $\ti < \tcl$, that is, when
the line of sight to the observer intersects the cloud volume, then
$\phi_+ = \pi$ when $\theta \leq \tcl - \ti$. Thus we have two cases,
as follows:
\begin{enumerate}
\item  $\ti < \tcl$
\begin{equation}
\phi_+ = \cases{\pi\;, & 
when $\ti - \tcl \leq \theta \leq \tcl - \ti \;,$ \cr\cr
\arccos \Theta\;, &
 when $\tcl -\ti \leq \theta \leq \ti + \tcl\;$ \cr\cr 0\; ,&
 otherwise; and \cr}\;
\end{equation}
\item $\ti > \tcl$
\begin{equation}
\phi_+ = \cases{\arccos \Theta\; ,&
 when $\ti -\tcl \leq \theta \leq \tcl + \ti\;,$ \cr\cr 0\; ,&
 otherwise, \cr}\;
\end{equation}
\end{enumerate}
where 
$$\Theta \equiv {\cos \tcl - \cos \theta \cos \ti\over \sin\theta\sin\ti }.$$

The stationary frame time $t_*$ is related to the observer time $t$
through 
\begin{equation}
t_z \equiv {t\over (1+z)} = t_*  -  {x\cos\theta\over c}\;,
\label{tz}
\end{equation}
where the zero of time in the two frames is defined by the start
of the GRB explosion. The differential distance $dx$ traveled by a
relativistic blast wave during the differential time elements $dt_*$,
$d\tp$ and $dt_x$ in the stationary, comoving, and observer frames,
respectively, is given by
\begin{equation}
dx = \beta c dt_* = \beta\Gamma c d\tp = \beta\Gamma\Dop c dt/(1+z)\; .
\label{dx}
\end{equation}
Eqs.\ (\ref{tz}) and (\ref{dx}) are needed to determine the
emissivity at assigned values of $x$ and $t_z$.

\subsection{Phases of the Interaction Event}

\begin{figure}
\epsscale{0.75}
\plotone{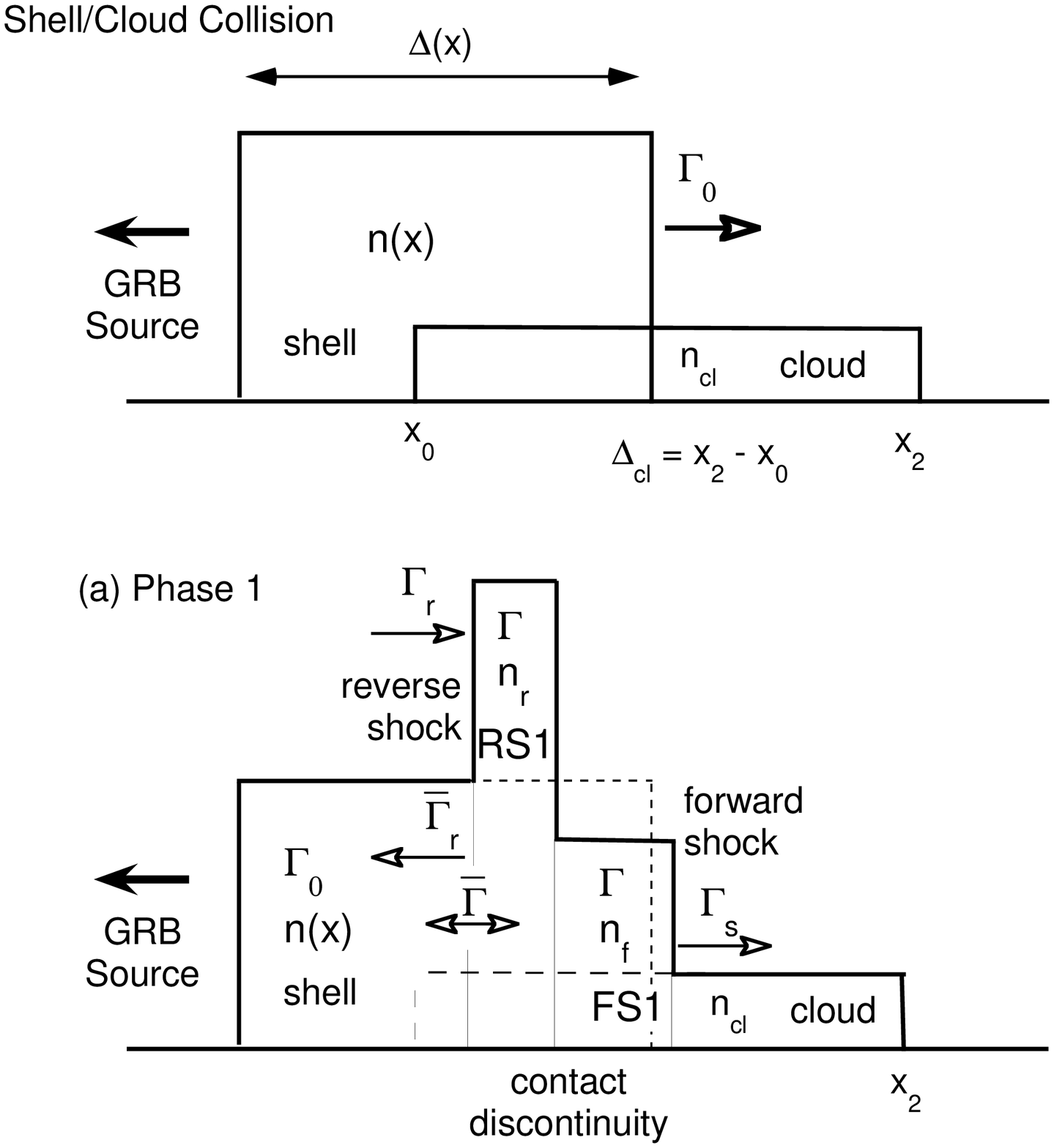}
\vskip-.60in
\plotone{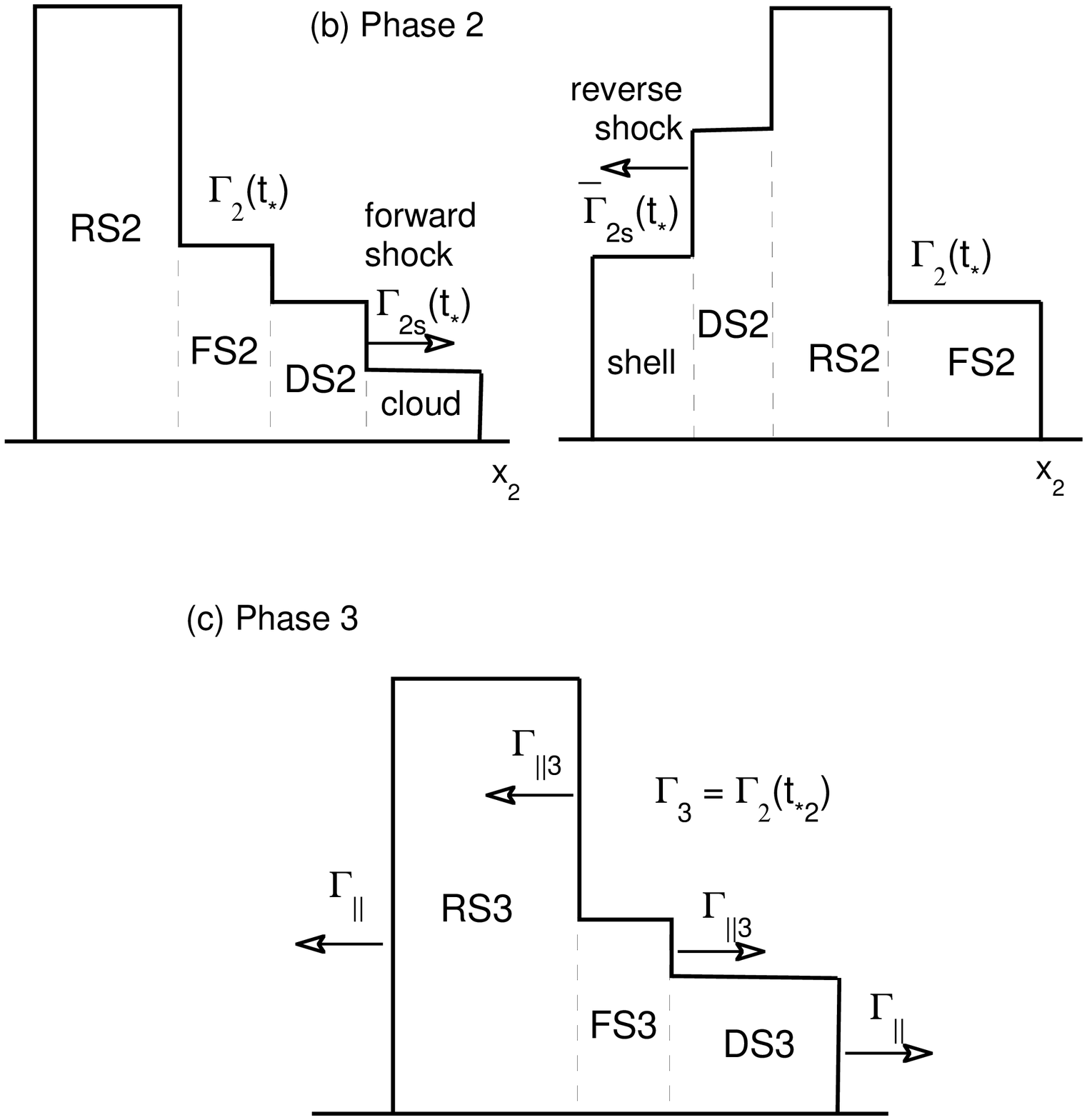}
\vskip-0.30in
\caption{ Phases of the Interaction Event. Top figure 
illustrates the planar geometry and widths, in the stationary frame,
of the shell and cloud. Panel (a) illustrates the Collision Phase,
Phase 1, with both forward shock (FS) and reverse shock (RS). Panel
(b) illustrates the Penetration Phase, Phase 2, when either the RS has
traversed the shell and the FS continues to pass through the cloud
(left), or when the FS has traversed the cloud and the RS continues to
pass through the shell (right).  Panel (c) illustrates the Expansion
Phase, Phase 3, where the shock fluid expands and the nonthermal
particles cool, for the former (left) case of Phase 2.}
\label{f2}
\end{figure}

The complete interaction is treated in a planar geometry, as
illustrated in Figs.\ 1 and 2. The cloud is assumed to have uniform
density between radii $x_0$ and $x_2$, so that the cloud width
$\Delta_{cl} = x_2 - x_0$.  A major simplifying assumption of the
analysis is that $\Delta_{cl} \ll x_0$, so that $n(x) \approx n(x_0)$
throughout the duration of the interaction. Three phases of the
interaction event are identified and illustrated in Fig.\ 2.

\begin{enumerate} 
\item Collision Phase (Fig.\ 2a). 
Both a forward shock (FS) and reverse shock (RS) are found in this
phase; the forward shock accelerates the cloud material, and the
reverse shock decelerates the shell material.
\item Penetration Phase (Fig.\ 2b). 
This phase bifurcates into two cases, depending on whether the RS
crosses the shell before the FS crosses the cloud, or whether the FS
crosses the cloud before the RS crosses the shell.  In the former
case, the shocked fluid produced during the collision phase
decelerates as more cloud material is swept up at the FS, during which
time a new accelerated particle population is introduced at the
decelerating forward shock, so we denote this process the deceleration
shock (DS). In the latter case, the shocked fluid produced during the
collision phase is accelerated by the remaining shell material at the
RS, and a new particle population is accelerated at the RS. We only
treat the first case of this phase (which is the most important case
for GRB studies) in this paper.
\item Expansion Phase (Fig.\ 2c). 
After the FS crosses the remaining cloud material in the first case of
Phase 2, or the RS crosses the remaining shell material in the second
case of Phase 2, the shocked fluid, being composed of highly
relativistic particles and magnetic fields, expands and adiabatically
cools.  \end{enumerate}

\subsubsection{Collision Phase}

We follow the approach of \cite{sp95} and \cite{kps99} to analyze this
phase.  Let $\bar\Gamma = 1/ \sqrt{1-\bar\beta^2}$ represent the
relative Lorentz factor of unshocked shell material in the rest frame
of the shocked fluid.  From Fig.\ 2b,
\begin{equation}
\bar\Gamma = \Gamma\Gamma_0(1-\beta\beta_0)\cong {1\over
2}({\Gamma\over\Gamma_0} + {\Gamma_0\over \Gamma})\;,
\label{Gammabar}
\end{equation} 
where the last relation applies in the regime $\Gamma_0,\Gamma \gg 1$
considered here.   Note that $\Gamma$ and $\bar\Gamma$ remain
constant during the duration of Phase 1, as a consequence of the
assumption that $\Delta_{cl} \ll x$.

The shocked fluid travels with Lorentz factor $\Gamma$, and the shock
itself moves with Lorentz factor $\Gamma_s$. When $\Gamma \gg 1$, then
$\Gamma_s \cong \sqrt{2}\Gamma$, implying a compression ratio of
$4\Gamma$ (particle densities will always be referred to the proper
frame). Likewise, when $\bar\Gamma \gg 1$, the Lorentz factor
of the RS in the comoving fluid frame is $\bar\Gamma_r \cong
\sqrt{2}\bar\Gamma$.  When $\bar\Gamma - 1 \ll 1$, then $\bar\beta_r =
\sqrt{1-\bar\Gamma_r^{-2}} \cong 4\bar\beta/3$, implying a compression
ratio $\cong 4$. Thus $\bar\Gamma_r \cong
\sqrt{2}\bar\Gamma$ when $\bar\Gamma > \sqrt{2}$, and $\bar\beta_r
\cong 4\bar\beta/3$ when $\bar\Gamma \leq \sqrt{2}$. The Lorentz factor of
the reverse shock in the stationary frame is $\Gamma_r =
\Gamma\bar\Gamma_r(1-\beta\bar\beta_r)$, and $\beta_r =
\sqrt{1-\Gamma_r^{-2}}$.

When $\Gamma \gg 1$, the fluid density of the FS material is $n_f\cong
4\Gamma n_{cl}$. The density of the reverse-shocked fluid is $n_r
\cong (4\bar\Gamma +3) n(x)$ for a relativistic reverse shock, and
$n_r \cong (4+5\bar\beta^2/4) n(x)$ for a nonrelativistic reverse
shock composed of cold shell material; an expression that joins the
two regimes is $n_r \approx 4\bar\Gamma n(x)$. The equality of
kinetic-energy densities at the contact discontinuity implies that
\begin{equation} 
 n_f(\Gamma - 1) = n_r (\bar\Gamma - 1 )\;\cong \; 4
n_{cl}\Gamma^2\; \cong \; 4n(x)(\bar\Gamma^2 -\bar\Gamma).
\label{kedensity}
\end{equation} 
The relativistic shock jump conditions for an isotropic explosion in a
uniform CBM therefore give \citep{sp95,pm99}
$${n(x_0)\over n_{cl}} \equiv F = {E_0\over 4\pi x_0^2 \Gamma_0^2 n_{cl}
m_pc^2
\Delta(x_0)}\cong$$
\begin{equation}
 {\Gamma^2\over
\bar\Gamma^2-\bar\Gamma}\rightarrow \cases{2\Gamma^2/\bar\beta^2\; ,&
${\rm ~~Nonrelativistic ~RS ~(NRS)}$ \cr\cr \Gamma^2/
\bar\Gamma^2\; ,& ${\rm ~~Relativistic ~RS ~(RRS)}$\cr}\;.
\label{n(x)/n}
\end{equation}
Hence
\begin{equation}
\bar\beta \cong \Gamma_0\sqrt{{2\over F}}\;,\; 
\bar \Gamma = {1\over \sqrt{1-\bar\beta^2}}\;,\;
\Gamma \cong {\Gamma_0\over(1+\bar\beta )}\;, \label{NRSap}
\end{equation}
$$\;{\rm when}\;F 
\gg 4\Gamma_0^2\; {\rm (NRS)},$$
and 
\begin{equation}
\bar\beta \cong 1 \;,\;\bar\Gamma = {\Gamma\over \sqrt{F}}\;,\;
\Gamma = {\Gamma_0 F^{1/4}\over \sqrt{2\Gamma_0 -F^{1/2}}}
\cong \sqrt{{\Gamma_0\over 2}} F^{1/4}\;,\; 
\label{RRSap}
\end{equation}
$$ 
{\rm when}\;F \ll 4\Gamma_0^2 \;{\rm (RRS)}\;.$$

\begin{figure}[t]
\epsscale{1.0}
\plotone{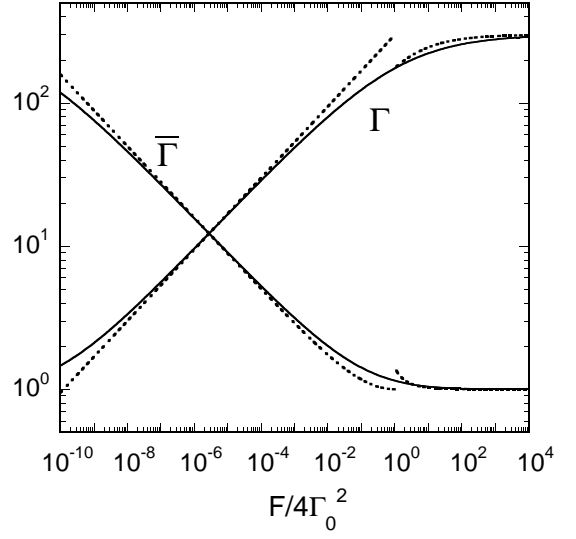}
\caption{ Lorentz factor $\Gamma$ of the shocked fluid in the explosion frame, 
and Lorentz factor $\bar\Gamma$ of the unshocked shell material in the
rest frame of the shocked fluid, as a function of $F/4\Gamma_0^2$ when
$\Gamma_0 = 300$. Solid curves give numerical results, and dotted
curves show approximations given in eqs.\ (\ref{NRSap}) and
(\ref{RRSap}).  }
\label{f3}
\end{figure}

Fig.\ 3 shows the shocked fluid Lorentz factor $\Gamma$ and the
relative Lorentz factor $\bar \Gamma$ as a function of $F/4\Gamma_0^2$
for $\Gamma_0 = 300$. The accurate numerical results are shown by the
solid curves, obtained by solving $F = \beta(\G^2
-\G)/\bar\beta(\bar\G^2 -\bar\G)$.  The dotted curves show the
approximate analytic expressions, eqs.\ (\ref{NRSap}) and
(\ref{RRSap}), using the expression $\Gamma \cong
\sqrt{\G_0/2}F^{1/4}$ in the latter case.  The numerical solutions for
$\Gamma$ and $\bar\Gamma$ are used in subsequent calculations.

The FS power is $dE^\prime/dt^\prime|_{FS} = A_{cl}n_{cl}m_pc^3
\beta(\Gamma^2 -\Gamma)\cong A_{cl}n_{cl}m_pc^3 \Gamma^2$.  The
reverse shock power is $dE^\prime/dt^\prime|_{RS} = A_{cl}
n(x_0)m_pc^3 \bar\beta(\bar\Gamma^2 -\bar\Gamma)$. Thus
$(dE^\prime/dt^\prime|_{RS})/(dE^\prime/dt^\prime|_{FS}) = 1$,
so that equal power is dissipated as internal energy in the FS
and the RS. The comoving duration of the FS
before penetrating the cloud is $\Delta \tp_{FS} \cong
\Delta_{cl}/\Gamma c$. The comoving duration of the RS before
penetrating the shell is $\Delta \tp_{RS} \approx
 \Gamma_0 \Delta/\bar\beta
\bar\Gamma c$.
Thus case (1), where  
the RS crosses the shell before the FS crosses the cloud, applies when
$\Delta \tp_{RS}/\Delta \tp_{FS} < 1$, that is, when
$\Delta/\Delta_{cl} < \bar\beta \bar\Gamma/\Gamma_0\Gamma\lesssim
1/\Gamma_0$. Case (1) generally applies when $\Delta \ll x/\Gamma_0^2$ (eq.\
[\ref{Delta(x)}]) and $\Delta_{cl} \lesssim x/\G_0$, which is
the condition for STV.

The range of $x$ occupied by the FS fluid during phase 1 (designated
as FS1) at time $t_* = t_z + x\cos\theta/c$ is
\begin{equation}
x_0 + \beta c(t_* - t_{*0}) \leq x \leq x_0 + \beta_{s} c(t_* - t_{*0})\;,
\label{x0FS1}
\end{equation}
while the range occupied by the RS fluid during phase 1, or RS1, is
\begin{equation}
x_0 + \beta_r c(t_* - t_{*0}) \leq x \leq x_0 + \beta c(t_* - t_{*0})\;,
\label{x0RS1}
\end{equation}
where $t_{*0} \equiv x_0 /\beta_0 c$ and $\beta_{s} = \sqrt{1-
\Gamma^{-2}_{s}}$.  The comoving time $\tp = (t_{*0}/\G_0) +(t_* -
t_{*0})/\Gamma$, so that the elapsed comoving time since the start of
the interaction is $\Delta\tp = (t_* - t_{*0})/\Gamma$.  In the
absence of an interaction, the shell occupies the range $\beta_0 c t_*
- \Delta \leq x \leq \beta_0 c t_*$. Thus the RS leaves the shell when
$\beta_0 c t_* - \Delta = x_0 + \beta_r c(t_* - t_{*0})$, that is,
when
\begin{equation}
t_* = t_{*\Delta} = t_{*0} + {\Delta\over (\beta_0 -\beta_r)c }\;.
\label{tstarDelta}
\end{equation}
This represents the end of Stage 1 for the case where the RS crosses
the shell before the FS crosses the cloud.  Therefore Stage 1 lasts
for comoving time
\begin{equation}
\Delta\tp_{RS} = {\Delta\over \Gamma(\beta_0 -\beta_r)c }\;,
\label{tprimeDelta}
\end{equation}
which may be compared with the approximation previously derived. The
time measured between the start of the GRB explosion and when the
blast wave first encounters the cloud is
\begin{equation}
t_0 = (1+z) {x_0\over \beta_0 c}\times\; 
 \, \cases{ 1-\beta_0\; & for $\ti\leq\tcl\;,$ \cr\cr
	  1-\beta_0\cos(\ti -\tcl ) & for $\ti > \tcl\;$.\cr} 
\label{tz0}
\end{equation}

Following the treatment of \cite{spn98} \cite[for the accuracy of this
approach, see][]{dcm00}, we let $n(\gamma;x,\tp)d\g$ represent
the differential number density of electrons with Lorentz factors
between $\gamma$ and $\gamma + d\gamma$ in the shocked fluid. Assuming
that all swept-up electrons are accelerated by the FS with spectral
index $p_f$, joint normalization of number and power gives a minimum
electron Lorentz factor
\begin{equation} \g_{min,f} \cong \epsilon_{e,f} {m_p\over m_e} 
f(p_f) \;(\Gamma -1)\;,\;{\rm ~where~} f(p) \equiv \;\big({p-2\over
p-1}\big)\;
\label{gminf}
\end{equation} 
and $\e_{e,f}$ represents the fraction of swept-up FS power carried by
the injected nonthermal electrons (subscript ``$f$" refers to the FS,
and subscript ``$r$" to the RS).  For electrons accelerated by the RS
with spectral index $p_r$, similar considerations give a minimum
Lorentz factor
\begin{equation} 
\g_{min,r} \cong \epsilon_{e,r} {m_p\over m_e}
f(p_r) \;(\bar\Gamma -1)\;.  
\label{gminfr} 
\end{equation} 

Electrons cool through synchrotron losses 
during comoving time $\Delta\tp$ since the start of the
collision to cooling Lorentz factor 
\begin{equation} 
\gamma_{c,f} = {6\pi m_e c\over \sigma_{\rm T}
B_{f}^2 \Delta\tp}\;\;{\rm and}\;\; \gamma_{c,r} = {6\pi m_e c\over
\sigma_{\rm T} B_{r}^2 \Delta\tp}\; 
\label{gammac} 
\end{equation} 
for the FS and RS fluids, respectively, assuming dominant synchrotron
losses. The magnetic field strength $B$ in the FS and RS fluids is
assigned according to the usual prescription, namely that the
magnetic-field energy density is proportional to the downstream energy
density of the shocked fluid with proportionality constant $\epsilon_B
\lesssim 1$. Thus
$$B_f^2 = 32 \pi n_{cl} m_pc^2 \epsilon_{B,f}
\beta(\Gamma^2 -\Gamma) \;,\; {\rm and}$$
\begin{equation}\;\; B_r^2 = 
32 \pi n(x) m_pc^2 \epsilon_{B,r}
\bar\beta(\bar\Gamma^2 -\bar\Gamma)\;.
\label{BfBr} 
\end{equation} 

The maximum electron Lorentz factor $\gamma_{max}$ is determined by
equating the most rapid acceleration rate expected in Fermi processes,
$\dot\g_{acc} = \e_2 eB/m_e c$, where the rate factor $\e_2 \lesssim
1$, with the synchrotron loss rate $|-\dot \g_{syn}|$, giving
\begin{equation} 
\g_{max,f} \cong ({6\pi e \e_{2,f}\over \sT B_f})^{1/2} 
\cong {1.2\times 10^8 \e_{2,f}\over \sqrt{B_f({\rm G})}}\;,
\label{g2} 
\end{equation} 
and similarly for $\g_{max,r}$ at the RS.

We calculate the nonthermal synchrotron radiation in the collision
phase spectrum by noting that the emissivity in eq.\ (\ref{fet1}) can
be expressed through the relation
\begin{equation}
\e^\prime j^\prime(\ep,\vec\Op;x,\tp) 
\cong {4\over 3} c\sT U_B \g_s^3 n(\g_s;x,\tp )\;,
\label{eprimejprime}
\end{equation}
where $U_B = B^2/8\pi$ is the magnetic-field energy density,
\begin{equation}
\g_s \cong \sqrt{\e_z\over \Dop(B/B_{cr})}\;,
\label{gammasyn}
\end{equation}
and the critical magnetic field $B_{cr} = m_e^2 c^3 /e\hbar =
4.414\times 10^{13}$ G.  This follows from the formula for
the electron energy-loss rate through synchrotron emission, and from 
the expression for the
mean energy of synchrotron photons radiated by an electron with
Lorentz factor $\g_s$. Eq.\ (\ref{eprimejprime}) applies to the FS and
RS fluids by using values of $B$ and $\g_s$ appropriate to each
component.

Assuming that the electrons are accelerated by the first-order Fermi
shock process and injected into the shocked fluid in the form of a
power law with index $p$, then the density distribution $n
(\gamma;x,\tp )$ of energized electrons can be approximated
by the expression
$$\g n(\g;x,\tp ) \cong $$
\begin{equation} 
 {n(x,\tp)\over s-1}\; 
 \, \cases{ (\g/\gamma_0)^{s-1}\; & for $\gamma_0
\leq\gamma\leq\gamma_{1}\;,$ \cr\cr
	 (\g_1/\g_0)^{1-s}(\g/\g_1)^{-p}\;, & for 
$\gamma_1 \leq \gamma \leq \gamma_{max}$.\cr} 
\label{gndist} 
\end{equation} 
In the slow-cooling regime, $\gamma_{min} < \gamma_c$, $\gamma_0 =
\gamma_{min}$, $\gamma_1 = \gamma_c$, and $s=p$.  In the fast-cooling
regime, $\gamma_c < \gamma_{min}$, $\gamma_0 = \gamma_c$, $\gamma_1 =
\gamma_{min}$, and $s = 2$ \citep{spn98}. At energies $\e \leq \e_0 =
(B/B_{cr})\g_0^2 \Dop/(1+z)$, corresponding to $\g \leq \g_0$, the
synchrotron spectrum is approximated by the elementary synchrotron
emissivity spectrum ($\propto \e^{1/3}$ in a $\nu F_\nu$ 
representation). Thus we can include this branch of the
synchrotron spectrum by writing $$\gamma n (\gamma;x,\tp )
\cong {n(x,\tp)\over s-1}\;({\e\over \e_0})^{1/3}\;{\rm ~for}\;
\gamma\leq \g_0\;.$$
Synchrotron self-absorption is neglected here, which
is valid for radiations detected at infrared frequencies and higher
considered in this paper.

The proper densities of electrons swept up by the FS and RS are
\begin{equation}
n_f (x,\tp) = { n_{cl}\over (\beta_s - \beta)} \cong 4\Gamma
n_{cl}\;\;{\rm and}\;\; n_r (x,\tp) = { \bar\beta \bar\Gamma n(x)\over
(\beta - \beta_r)\Gamma^2}\;,
\label{Nftp}
\end{equation}
respectively.  In the treatment used here, the total electron
density is independent of location within the volume of the FS and RS
fluids during Phase 1.  The collision phase spectrum is calculated by
substituting eq.\ (\ref{eprimejprime}) into eq.\ (\ref{fet2}), making
use of eq.\ (\ref{gammasyn}) to relate $\g$ to the received photon
energy.

\subsubsection{Penetration Phase}

Only the case where the RS passes through the shell before the FS
passes through the cloud is analyzed here.  During this phase, the
shell is assumed to be hydrodynamically connected so that the entire
shocked fluid decelerates with the same Lorentz factor $\G_2$. As the
FS passes through the remaining cloud material, it sweeps up and
injects a new nonthermal particle population in addition to the
nonthermal electrons left over from Phase 1. We denote $\Gamma =
\Gamma_1$, the Lorentz factor of the shocked fluid during Phase 1. The
equation for blast-wave deceleration of a relativistic blast wave when
it sweeps up material from the cloud is given by
\begin{equation}
\G_2(x) = {\G_1 \over [ 1 + 
(2-\varphi)\G_1^2 c^2 m(x)/E_0]^{1/(2-\varphi)}}\;
\label{G2}
\end{equation}
 \citep{bd00,dh01}, where $\varphi$ is the fraction of swept-up energy that
 is radiated, and $m(x) = 4\pi m_p \int_{x_\Delta}^x \;dx^\prime
 x^{\prime 2} n_{cl}(x^\prime )
\cong 4\pi m_p n_{cl} \xD^2(x-\xD)$
is the swept-up mass. Here 
\begin{equation}
\xD = x_0 + {\beta_s \Delta \over \beta_0 -\beta_r}
\label{xDelta}
\end{equation}
is the distance of the FS from the explosion center at the end of
Phase 1 (see eqs.\ [\ref{x0FS1}] and [\ref{tstarDelta}]).  Defining
the deceleration radius $x_d$ through the expression $\G_1 m(x_d) =
E_0/\G_1 c^2$ gives the following expression for the deceleration
radius for a blast-wave/cloud interaction:
\begin{equation}
x_d = \xD + { E_0\over 4\pi m_pc^2 \G_1^2 n_{cl} \xD^2}\;.
\label{xdec}
\end{equation}
Assuming adiabatic evolution ($\varphi \ll 1$), eq.\ (\ref{G2}) becomes 
\begin{equation}
\G_2 = {\G_1 \over \sqrt{1+2({x-\xD\over x_d -\xD})}} = 
{\G_1 \over \sqrt{1+2\beta_{1s}u - u^2/2\G_1^2}}\;,
\label{G2xt}
\end{equation}
where $u \equiv c(t_* - t_{*0})/(x_d - \xD )$, and $\beta_{1s} =
\sqrt{1-1/2\G_1^2}$. The expression on the right-hand-side of eq.\
(\ref{G2xt}) is obtained by integrating $dx = \beta_{2s}(x) dt_*$,
where $\G_{2s} (x) = \sqrt{2} \Gamma_2(x)$.

During Phase 2, the particles injected at the RS and FS during Phase 1
are no longer subject to further acceleration. We assume that these
fluid volumes, denoted RS2 and FS2, respectively, neither expand nor
contract during Phase 2, and decelerate with Lorentz factor
$\G_2(t_*)$.  The range occupied by RS2 is therefore
\begin{equation}
x_0 + {\beta_r \Delta \over \beta_0 - \beta_r} + x_2(t_* -
t_{*\Delta}) \leq x \leq x_0 + {\beta_1 \Delta \over \beta_0 -
\beta_r} + x_2(t_* - t_{*\Delta})\;.
\label{RS2range}
\end{equation}
For FS2, the range is 
\begin{equation}
x_0 + {\beta_1 \Delta \over \beta_0 - \beta_r} + x_2(t_* -
t_{*\Delta}) \leq x \leq x_\Delta + x_2(t_* - t_{*\Delta})\;.
\label{FS2range}
\end{equation}
Here 
$$x_2(t_* - t_{*\Delta}) = c\int_{t_{*\Delta}}^{t_*}dt^\prime_*
\;\beta_2(t^\prime_*) = $$
$${(x_d - x_\Delta)\over \Gamma_1} \big[\;\big(
{u^\prime\over 2} - \beta_{1s}\G_1^2\big) \sqrt{ {u^{\prime 2}\over
2\G_1^2} - 2\beta_{1s}u^\prime+\G_1^2 -1 }$$
\begin{equation}
- {\G_1^3 \over \sqrt{2}}\;\ln | {\sqrt{2}\over \G_1}
\sqrt{{u^{\prime 2}\over 2\G_1^2} - 
2\beta_{1s}u^\prime+\G_1^2 -1 } +{u^\prime\over \G_1^2} - 2\beta_{1s} |
\;\big]\big|_0^u\;.
\label{FS2range1}
\end{equation}

The elapsed comoving time since the start of Phase 2, obtained by
integrating $d\tp = dt_*/\G_2(t_* )$, is given by
$$\tp -\tp_\Delta = {x_d - \xD\over c\G_1}
\{({u\over 2} - \beta_{1s}\G_1^2)\sqrt{1+\beta_{1s}u - {u^2\over 2\G_1^2}}
+ $$
\begin{equation} 
\beta_{1s}\G_1^2 - \sqrt{2}\G_1^3 [\arcsin (\beta_{1s}- {u \over
2\G_1^2 }) - \arcsin \beta_{1s} ]\}\;.
\label{tpPhase2}
\end{equation}
The RS2 and FS2 emissivities are the same as the RS1 and FS1
emissivities, except now $\Delta \tp = \tp -\tp_\Delta + (t_{*\Delta}
- t_0)/\Gamma_1$ is used to evaluate the cooling Lorentz factor, eq.\
(\ref{gammac}). Moreover, because injection and subsequent particle
acceleration has ceased in these two phases, the maximum electron
Lorentz factor decays by synchrotron losses to a value of
\begin{equation}
\gamma_{2,f}(\tp ) = [ \gamma_{max,f}^{-1} + 
{\sigma_T B_f^2 \over 6\pi m_e c} (\tp -\tp_\Delta)]^{-1}\;,
\label{gamma2}
\end{equation}
where $\gamma_{max,f}$ is given by eq.\ (\ref{g2}) for FS2, with a
related equation for RS2. Because the FS and RS fluid shells are
assumed not to expand during Phase 2, the magnetic field remains the
same as in Phase 1.

The fluid containing the nonthermal electrons and protons injected by
the decelerating shock as it sweeps up cloud material in Phase 2,
denoted DS2, occupies the range
\begin{equation}
x_\Delta + x_2(t_* - t_{*\Delta}) \leq x \leq x_\Delta + x_{2s}(t_* -
t_{*\Delta})\;,
\label{DS2range}
\end{equation}
where $$x_{2s}(t_* - t_{*\Delta}) =
c\int_{t_{*\Delta}}^{t_*}dt^\prime_* \;\beta_{2s}(t^\prime_*) = {(x_d - x_\Delta)\over \sqrt{2}\Gamma_1}\times$$
\begin{equation}
  \big[\;\big( {u\over 2} -
\beta_{1s}\G_1^2\big) \sqrt{ {u^{ 2}\over 2\G_1^2} - 2\beta_{1s}u
+2\G_1^2 -1 } +\beta_{1s}\G_1^2 \sqrt{2\G_1^2 -1}\;\big]\;.
\label{x2s}
\end{equation}
The DS2 emissivity at time $t_*$, determined by the values of $\theta$,
$x$, and observer time $t_z$ in the evaluation of eq.\ (\ref{fet2}), is
produced by nonthermal electrons injected at location $x_i$ and time
$t_{*i}$ . The values of $x_i$ and $t_{*i}$ are determined by the
intersection of $x_{2s} = \xD + x_{2s}(t_{*i} - t_{*\Delta} )\;$,
representing the leading edge of the deceleration shock (DS) that
injects particles into the fluid, and $x_i = x -x_2 (t_* - t_{*i})$,
representing the worldline connecting the injection coordinates to the
received values. The injection coordinates $x_i$ and $u_i = c(t_{*i} -
t_{*\Delta})/(x_d - \xD)$ are therefore obtained by numerically
solving $x - \xD - x_2(t_* - t_{*\Delta}) = x_{2s}(t_{*i} -
t_{*\Delta}) - x_2(t_{*i} - t_{*\Delta})$ using a staggered leapfrog
routine.

The received spectrum during Phase 2 is given by eq.\ (\ref{fet2}),
with the $x$-integral limited to the ranges defined above. The
emissivity for DS2 is given by eq.\ (\ref{eprimejprime}), with values
of $\g_{min}$, $\g_c$ and $\g_{max}$ appropriate to the injection
value of $\Gamma(t_{*i})$.  The elapsed comoving time since injection
used to calculate $\g_c$ is obtained from eq.\ (\ref{tpPhase2}).  The
Doppler factor at time $t_*$ uses eq.\ (\ref{G2xt}) for $\Gamma$.
Phase 2 ends when $x = x_2$, that is, at stationary time 
$$t_{*2} =
t_{*\Delta} +$$
\begin{equation}
 {\sqrt{2} \G_1 (x_d - \xD )\over c} \big[\;\sqrt{2\G_1^2
-1 } - \sqrt{2\G_1^2 -1 -2 ({x_2 - \xD\over x_s - \xD})}\;\big]\;.
\label{e2star}
\end{equation}

\subsubsection{Expansion Phase} 

After the FS passes through the cloud, the shocked fluid has Lorentz
factor
\begin{equation}
\G_3 = \G_2(t_{*2}) = {\G_1\over \sqrt{1+ 
2({x_2 - x_\Delta\over x_d - x_\Delta})}}\;.
\label{G3}
\end{equation}
Because there is no further pressure on the fluid, the shocked fluid
is assumed to expand outward along the parallel direction with speed
$\beta_\parallel c$ and Lorentz factor 
$$\G_\parallel =
1/\sqrt{1-\beta_\parallel^2}\;.$$  (Because the shocked fluid has a much
smaller parallel than transverse extent, due to the compression of the
quasi-spherical cloud material by a factor $\gtrsim 4\Gamma_3$ in the
parallel direction, effects of transverse expansion can be neglected.)
No further injection or acceleration takes place in Phase 3, and the
relativistic electrons cool by synchrotron and expansion losses.  The
shocked fluid in Phase 3, denoted by RS3, FS3, and DS3, consists of
fluid from the prior RS, FS, and DS phases.  In order that each fluid phase
is to have the same fractional volume expansion, the FS fluid is
assumed to expand in both transverse directions in the comoving frame
with speed $\beta_\parallel c/3$, where $\G_{\parallel 3} =
1/\sqrt{1-(\beta_\parallel/3)^2}$.

From these assumptions, we define the four stationary frame Lorentz
factors $\G_{RS3} = \G_3 \G_\parallel (1-\beta_3 \beta_\parallel
)\;,\; \G_{FS3-} = \G_3 \G_{\parallel 3} (1-\beta_3 \beta_{\parallel
3} )\;,\; \G_{FS3+} = \G_3 \G_{\parallel 3} (1+\beta_3
\beta_{\parallel 3} )\;{\rm and}\;
\G_{DS3} = \G_3 \G_{\parallel} (1+\beta_3 \beta_{\parallel} )\;,$ along
with their associated $\beta$ factors $\beta_{RS3},
\beta_{FS3-},\beta_{FS3+}$ and $\beta_{DS3}$, respectively, that
characterize the motions of the radial boundaries of the different
fluid layers. The range of RS3 at time $t_*$ is
$$x_0 + {\beta_r \Delta \over \beta_0 - \beta_r} +
x_2(t_{*2}- t_{*\Delta})+c\beta_{RS3}(t_* - t_{*2})
$$
\begin{equation}
 \leq x   \leq 
x_0 + {\beta_1 \Delta \over \beta_0 - \beta_r} + 
x_2(t_{*2}- t_{*\Delta})+c\beta_{FS3-}(t_* - t_{*2})\;.
\label{RS3range}
\end{equation}
For FS3, the range is 
$$x_0 + {\beta_1 \Delta \over \beta_0 - \beta_r} + 
x_2(t_{*2}- t_{*\Delta})+c\beta_{FS3-}(t_* - t_{*2})
$$
\begin{equation}
 \leq x \leq 
x_\Delta + x_2(t_{*2}- t_{*\Delta})+c\beta_{FS3+}
(t_* - t_{*2})\;.
\label{FS3range}
\end{equation}

The range of DS3 is 
$$x_\Delta + x_2(t_{*2}- t_{*\Delta})+ 
c\beta_{FS3+}(t_* - t_{*2})\leq x \leq 
$$
\begin{equation}
 x_\Delta + x_{2s}(t_{*2}- t_{*\Delta})+ 
c\beta_{DS3}(t_* - t_{*2})\;.
\label{DS3range}
\end{equation}

Let $R^\prime_\parallel (\tp ) = R^{\prime 0}_\parallel +
2\beta_\parallel c \tp$ denote the comoving parallel width of the
entire shocked fluid, with additional subscripts $i = $ ``RS," ``FS,"
and ``DS" to refer to the widths $R^\prime_{\parallel i} (\tp ) =
R^{\prime 0}_{\parallel i} + {2\over 3} \beta_\parallel c(\tp -\tp_2
)$ of the forward, reverse, and deceleration shocked fluid layers,
respectively. The superscript ``0" refers to the radial width of the
fluid layer at the end of Phase 2, and $\tp_2$ is the comoving time at
the end of Phase 2. From eqs.\ (\ref{RS3range}) -- (\ref{DS3range}),
$R^{\prime 0}_{\parallel RS} = \G_3 (\beta_1 -\beta_r)\Delta/(\beta_0
-\beta_r)$, $R^{\prime 0}_{\parallel FS} = \G_3 (\beta_s
-\beta_r)\Delta/(\beta_0 -\beta_r)$, and $R^{\prime 0}_{\parallel DS}
= \G_3 [x_{2s}(t_{*2}- t_{*\Delta}) - x_2(t_{*2}- t_{*\Delta})]$, with
$R^{\prime 0}_\parallel$ equal to the sum of these three widths.

Conservation of magnetic flux for the transverse magnetic-field
component $B^0_\perp$ implies $B_\perp R_\parallel^{\prime 2} = const
= B_\perp^0 R_\parallel^{\prime 0 ~2}$, so that $B_\perp(\tp ) =
B_\perp^0 R_\parallel^{\prime 0 ~2}/ [R^{\prime 0}_\parallel + {2\over
3} \beta_\parallel c(\tp -\tp_2 )]^2$. If the electrons are
isotropized on time scales short compared to the cooling timescale,
then the electron energy loss-rate due to adiabatic and synchrotron
losses is given by
\begin{equation}
-{d\gamma\over d\tp} = {1\over
R^\prime_\parallel}{dR^\prime_\parallel\over d\tp}\;\gamma +
{\sigma_{\rm T} [B_\perp^2(\tp ) + B_\parallel^2(\tp )]\over 6\pi m_e
c}~\gamma^2\;,
\label{dgammadtp}
\end{equation}
where $B_\parallel$ is the parallel magnetic-field component. Assuming
that $B_\perp(\tp )
\gg B_\parallel(\tp )$, which would occur 
if the magnetic field is formed by sweeping up
and compressing an external field, then eq.\ (\ref{dgammadtp}) can be
written in the form
\begin{equation}
-{d\gamma\over d\tau} = {\gamma\over\tau} +
b {\gamma^2\over \tau^4}\;,
\label{dgammadtau}
\end{equation}
where $\tau \equiv 1 + {2\over 3}\beta_\parallel c(\tp -\tp_2
)/R_\parallel^{\prime 0}$ and $b \equiv R_\parallel^{\prime
0}\sigma_{\rm T}B_\perp^{0~2}/ ( 4\pi \beta_\parallel m_ec^2)$.  This
can be solved \citep{gbd06} using the substitution $y = \gamma
\tau^{-n}$, with $n = 3$, giving the result
\begin{equation}
\gamma(\tau) = {4\tau^3\over b(\tau^4 - 1) + 4\tau^4/\gamma_i}\;,
\label{gammatau}
\end{equation}
where $\g_i$ is the initial electron Lorentz factor at $\tp = \tp_2$
or $\tau = 1$.

The emissivity for RS3 and FS3 is given by eq.\ (\ref{eprimejprime})
with
\begin{equation}
n(\gp;\tp) = n(\g;\tau) = 
[{R_\parallel(\tp )\over R_\parallel^{\prime ~0}}]^{-1}
{n(\g_i;1)\g_i^2 \over \tau\g^2} 
= {n(\g_i;1)\g_i^2 \over \tau^2\g^2}\;,
\label{nprimegammatp}
\end{equation}
$B \rightarrow B_\perp(\tp ) = B_\perp^0/\tau^2$, and
$n(\g_i;1)$ is the electron $\gamma$ distribution at the end of
Phase 2.  This expression also applies to the DS3 emissivity, with the
emission coordinates $x,t_*$ mapped back to the injection coordinates
$x_i,t_{*i}$ as was done for DS2.  Note the minor inconsistency for
Phase 3 between the original assumption that the magnetic field is
randomly ordered, compared to the dominant perpendicular magnetic
field component implied by field compression in this phase. This
inconsistency should be relaxed with the use of directional
synchrotron emissivity spectra in more detailed treatments but, in
either case, the contribution from the expansion phase is usually much
smaller than that from the penetration phase.  The Doppler factor
during Phase 3 is $\Dop = [\G_3(1-\beta_3\mu)]^{-1}$.

\section{Results}

Nonthermal synchrotron emission spectra from blast-wave/cloud
interactions were computed using the formulae given in the previous
section. Standard parameters are given in Table 1.  For the model GRB
pulse, we consider a source at $z = 1$, corresponding to the mean
redshift of the pre-Swift BATSE/Beppo-SAX/HETE-2/INTEGRAL sample. The
coasting Lorentz factor and apparent isotropic energy release of the
blast wave are chosen to be $\Gamma_0 = 300$ and $E_0 = 10^{53}$ ergs,
respectively.  The initial width of the shell is $\Delta_0 = 10^7$ cm.
The blast wave is assumed to encounter a small cloud with nominal
density $n_{cl} = 10^3$ cm$^{-3}$ located at a distance $x_0 =
10^{16}$ cm from the explosion center\footnote{The total mass of such
clouds distributed in all directions around the GRB source is
$\lesssim 10^{-5} M_\odot$.}  centered along the line-of-sight to the
observer.  To simulate a quasi-spherical cloud, the cloud width
$\Delta_{cl} = 2\times 10^{14}$ cm and $\theta_{cl} = 0.01$. Other
standard parameters are injection electron index $p = 2.5$,
$\epsilon_e = 0.1$, and $\epsilon_2 = 0.1$. The parameters
$\e_e,\e_B,$ $\e_2$, and $p$ of the FS and RS are assumed to be the
same throughout all phases of the interaction in the calculations
presented here.

\begin{figure}[t]
\epsscale{1.0}
\plotone{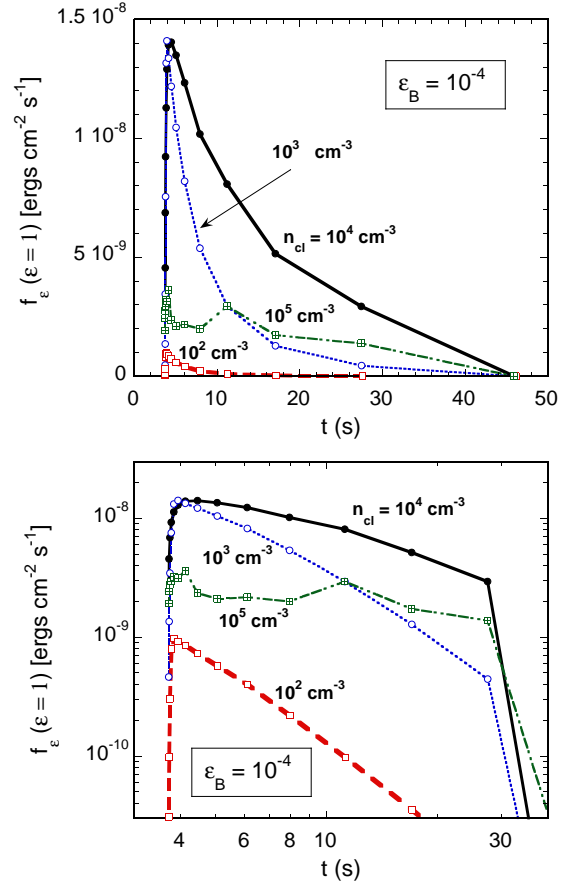}
\caption{Lightcurves calculated at 511 keV for a standard 
model GRB pulse with $\e_B = 10^{-4}$ and $\eta = 1/\Gamma_0$.  The
cloud geometry is fixed with $\theta_{cl} = 3/\Gamma_0$, but its
density is varied as labeled by the curves. Top figure: linear scale.
Bottom figure: logarithmic scale.}
\label{f4}
\end{figure}

Fig.\ 4 shows lightcurves computed at 511 keV for the standard model
GRB pulse with $\epsilon_B = 10^{-4}$. We prove in Section 4.2 that a
rapidly variable bright pulse cannot be made when $\eta = 1$, and
numerical simulations with $\eta = 1$ confirm this result. Thus we
adopt the thin-shell assumption ($\eta = 1/\Gamma_0$) in Fig.\ 4, and
assume that the cloud is on the axis to the line-of-sight to the
observer. This calculation shows the dependence of the behavior of the
light curve on cloud density $n_{cl}$.

The kinematic minimum and maximum times are implied by the idealized
cloud location and geometry. From eq.\ (\ref{tz0}) for on-axis clouds,
$$t_{min} \equiv {(1+z) x_0\over 2\Gamma_0^2 c}\cong
 3.70\; 
\left({1+z\over 2}\right)\;{x_{16}\over \Gamma_{300}^2}\;{\rm s}\;{\rm~and~}\;$$
\begin{equation}
 t_{max} \approx t_{min}(1+\Gamma_0^2\theta_{cl}^2)\;;
\label{tmin}
\end{equation}
here $\Gamma_{300} = \Gamma_0/300$. 
Depending on observing energy, the maximum emission time will also be
extended by the timescale for electron cooling.  As can be seen from
Fig.\ 4, at low densities the flux is weak and the blast wave hardly
decelerates.  For most purposes, such interactions can be neglected as
they extract only a small amount of energy from that solid angle
element of the blast wave. With increasing $n_{cl}$, the $\nu F_\nu$
flux increases until the cloud density becomes so thick that the RS
becomes relativistic, $\Gamma \ll \Gamma_0$, and the emission is
received at lower photon energies. At such high densities, the flux
received by a distant observer is weak but may last for a long time,
especially at small photon energies.

The model light curves show a sharp rise followed by a more leisurely
power-law temporal decay.  Some of this rapid rise has to do with the
idealized cloud geometry: the use of an actual spherical cloud
geometry (which could be mimicked by superposition of interacting
annuli) would tend to reduce the sharpness in the rising phase of the
GRB light curve.  As can be seen from the bottom figure of Fig.\ 4 in a
logarithmic representation, the power-law decay of the flux displays a
temporal curvature or softening at times well past the peak from
cooling breaks in the synchrotron spectrum.  This is related to the
detection of higher energy photons that were emitted from a
higher-energy and usually softer part of the spectrum by off-axis
emitting sites
\citep{fmn96,kp00,der04,lia06,zha07}.  The pulse shapes with $n_{cl} \ll
10^{4}$ cm$^{-3}$ correspond to kinematic ``curvature" pulses
\citep{der04}, where the radiating shell is thin and the emission is
radiated promptly on the crossing time defined by the comoving shell
thickness, and the spectrum is well described by a single power law in
the weak cooling regime.  The GRB light curve produced by interactions
with such clouds with optimal density $n_{cl} \cong 10^3$ -- $10^5$
cm$^{-3}$ would represent a generic long-duration GRB except that it
is rather weak, reaching peak $\nu F_\nu$ fluxes of $\sim 1$ --
$2\times 10^{-8}$ ergs cm$^{-2}$ s$^{-1}$ at $\sim 500$
keV.\footnote{The $\nu F_\nu$ peak flux measured by a GRB telescope
integrating over a finite bandwidth would be a factor of a few larger
due to a bolometric spectral correction.}

\begin{figure}[t]
\vskip-2.5in
\epsscale{1.2}
\plotone{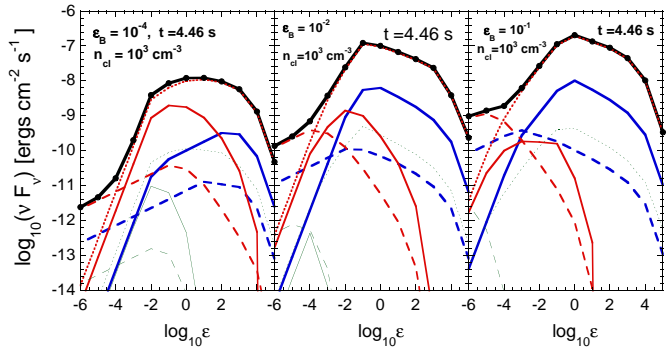}
\caption{ Separate spectral components and the total $\nu F_\nu$ flux
observed at $4.46$ s after the start of the GRB, near
the peak flux of its lightcurve. This GRB pulse model uses parameters
given in Table 1, with $n_{cl} = 10^3$ cm$^{-3}$ and $\eta =
1/\Gamma_0$. The differences between the system are that, from left to
right, $\e_B = 10^{-4}$, $\e_B = 10^{-2}$, and $\e_B = 0.1$.  The
total spectrum is given by the heavy solid curve with data points, the
FS components are given by the solid curves, the RS components by the
dashed curves, and the DS components by the dotted curves. The heavy
blue, medium red, and light green curves correspond to Phases 1, 2 and
3, respectively. }
\label{f5}
\end{figure}

The weakness of the flux is largely due to the small value used for
the $\e_B$ parameter. Fig.\ 5 shows the $\nu F_\nu$ SEDs calculated
near the peak of the $\e = 1$ light curve at observer time $t = 4.46$
s, for three different cases of $\e_B$ and with $n_{cl} = 10^3$
cm$^{-3}$.  Other parameters of the event are the same as in Fig.\ 4.
The various components making up the spectrum are shown.  The DS
component during Phase 2 makes the dominant contribution to the $\nu
F_\nu$ spectra near peak energy release. For the higher magnetic field
case $\e_B = 0.01$, and even more so for $\e_B = 0.1$, the stronger
cooling causes the Collision Phase components to decay away more
rapidly.

For the case $\e_B = 10^{-4}$ on the left, one sees that the peak $\nu
F_\nu$ flux, which is important for triggering, is at the level of a
$few\times10^{-8}$ ergs cm$^{-2}$ s$^{-1}$. BATSE triggered between
$\e \cong 0.1$ -- $0.6$ at a typical $\nu F_\nu$ flux level of
$\approx 5\times 10^{-8}$ -- $10^{-7}$ ergs cm$^{-2}$ s$^{-1}$, so
such a GRB would be hard to detect unless it occurred nearby ($z
\lesssim 1$) and happenened to be pointed at us. This cannot be
excluded for GRBs without redshift measurements, so our interest for
spectral modeling is principally on GRBs that have redshift
information. Stronger magnetic fields produce brighter pulses, and
when $\e_B \gtrsim 0.01$, the model pulse flux is at the level
$\gtrsim 3\times 10^{-7}$ ergs cm$^{-2}$ s$^{-1}$, making it an easily
detectable GRB with BATSE at $z \cong 1$, and even more so when $\e_B
\gtrsim 0.1$ and $\e_e \gtrsim 0.1$, although the adiabatic assumption
for the deceleration shock may then become invalid, a situation that
can also arise even when $\e_e\sim 0.1$ if photohadronic energy losses
are strong \citep{der06}.

An interesting result of these calculations is that the spectral index
below the $\nu F_\nu$ peak is very hard, in number index harder than
$-0.7$, $-1$, and $-4/3$ for $\e_B = 10^{-4}, 10^{-2},$ and $10^{-1}$,
respectively. In no case does a well-defined cooling spectrum with
number index $-3/2$ or softer form. This suggests another argument
against the concerns of \citet{gcl00} that radiative cooling spectra
should be observed in GRB spectra. Here there is not sufficient time
until the interaction is over for the electrons to cool, so the
spectra remain much harder than a cooling synchrotron spectrum below
$E_{pk}$. What stops the interaction and cooling is simply that, after
both shocks have passed through the shell and cloud, the disturbed
material rapidly expands, shutting off the subsequent synchrotron
emissions as nonradiative adiabatic losses increase and the magnetic
field intensity, and therefore the synchrotron
losses, decrease.

The spectra formed by the interaction of a relativistic blast wave
with a small cloud consisting of both FS and RS emission have the
characteristic ``Band''-shape \citep{ban93}, except for a lower energy
emission component from the RSs.  The relative importance of the RS and
FS components increases with increasing $\e_B$ for systems with
nonrelativistic or weakly relativistic RS.  When $\epsilon_B \gg
10^{-4}$, the RS produces a strong optical flux.  Indeed, this is the
external-shock model explanation for prompt optical emission in GRB
emission, first discovered by \citet{ake99}.

Even when $\epsilon_B= 10^{-4}$, the RS emission produces concavity in
the X-ray spectrum that could be detectable in joint spectral fitting
of the Swift BAT and XRT data with optical/UV measurements with the
UVOT and ground-based robotic optical telescopes.  For GRBs as modeled
by Fig.\ 5, optical flashes as bright as $m_V \approx 10$ -- 17 would
be coincident with GRB pulses.  The relative amplitude of the reverse
shock component is also, however, very sensitive to the value of the
shock thickness parameter $\eta$, and various underlying assumptions
for the constancy of the $\e_e$ and $\e_B$ parameters with time and
equality in the reverse and forward shocks. The rich parameter space
could allow detailed modeling of prompt and early afterglow optical
light curves from, e.g., RAPTOR, ROTSE, and super-LOTIS.

\begin{figure}[t]
\epsscale{1.0}
\plotone{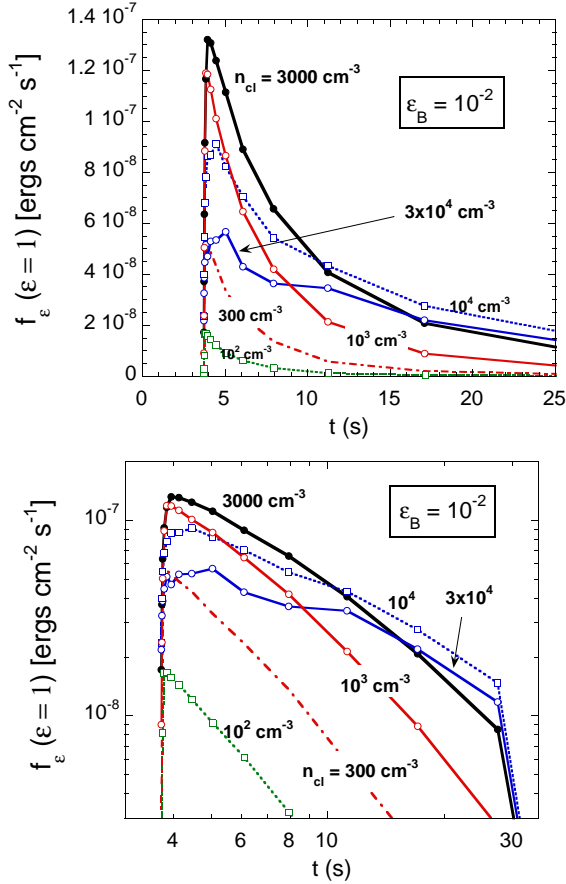}
\caption{ Light curves at $\epsilon = 1$ ($h\nu = 511$ keV) 
from the interaction of a GRB blast wave with a cloud for standard GRB
pulse model with $\e_B = 10^{-2}$, $\eta = 1/\G_0$, and other
parameters given in Table 1, for different values of $n_{cl}$.
Top figure: linear scale.
Bottom figure: logarithmic scale. }
\label{f6}
\end{figure}

Fig.\ 6 shows 511 keV light curves calculated for the standard GRB
pulse parameter set with $\epsilon_B = 10^{-2}$, while varying the
cloud density $n_{cl}$. The same basic behavior as was found in Fig.\
4 is apparent, except that the pulse is much brighter and decays more
rapidly. The optimal cloud density to make the brightest possible
pulse is again near $n_{cl} \cong 10^3$ cm$^{-3}$ for $\epsilon_B =
10^{-2}$ (see Section 4.2).  The generic pulse profile found here is
typical of a ``FRED"-type GRB.\footnote{FRED stands for Fast Rise,
Exponential Decay, though the decay phase is often better fit by a
power law.}  Such a profile could just as easily be produced by a GRB
blast wave decelerating in a uniform surrounding medium
\citep{dbc99}. The external shock model can explain relatively smooth
light curves that moreover show various widths and asymmetries. The
central question for GRB pulse modeling is, however, whether the rapid
variability found in some GRB light curves can be reproduced in an
external shock model.

\begin{figure}[t]
\epsscale{1.0}
\plotone{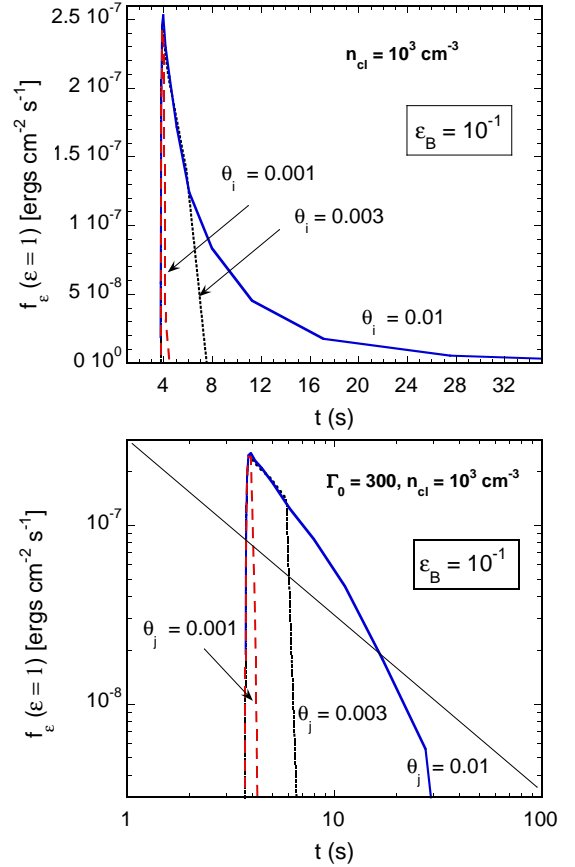}
\caption{ Calculated light curves at $\epsilon = 1$ ($h\nu = 511$ keV) 
from a GRB blast wave impacting a cloud with opening half-angle
$\theta_{cl} = 0.01$ located on the symmetry axis of the jet and the
line-of-sight to the observer. Parameters for the standard GRB pulse
model given in Table 1 are used, with $\e_B = 0.1$ and $\eta =
1/\Gamma_0$.  The extent of the jet opening angle is $\theta_{j} =
0.01, 0.003$, and $0.001$, as labeled. Top figure: linear scale.
Bottom figure: logarithmic scale. The straight line is $\propto
t^{-1}$. }
\label{f7}
\end{figure}

Fig.\ 7 displays the light curve formed for the standard 
parameter set with $\e_B = 10^{-1}$ and $\eta = 1/\Gamma_0$. 
Here the cloud has extent $\theta_{cl} = 0.01$, and a jet with 
opening half-angle $\theta_j$ collides with the cloud.
This plot indicates the latitudes from which the pulse
flux originates. The high-latitude emission arrives at late times, and
most of the fluence is formed by interactions for jet interactions
within the Doppler opening angle $1/\Gamma_0$.
Although this calculation illustrates the part of the blast-wave/cloud
interaction from where the flux originates,
the opposite limit, where $\theta_j \gg \theta_{cl}$, is of more 
interest for producing STV in GRB light curves. 

\begin{figure}[t]
\epsscale{1.0}
\plotone{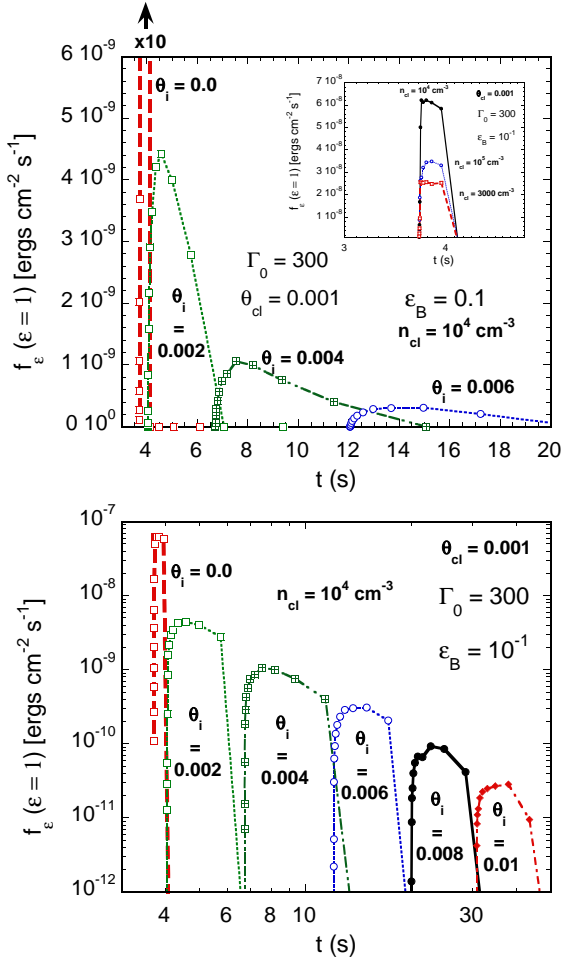}
\caption{Pulses formed by GRB blast wave shells with $\eta = 1/\Gamma_0$ 
colliding with clouds at different
inclinations from the line of sight.  The clouds have
 half-opening angle $\theta_{cl} = 0.001$ which, for $\Gamma_0 =
300$, represents an angular extent equal to $1/3$ of the Doppler cone.
Top figure: linear scale. Inset shows
detail of brightest pulse. Bottom figure: logarithmic scale. }
\label{f8}
\end{figure}

 Fig.\ 8 shows the emission profiles produced by clouds with density
 $n_{cl} = 10^4$ cm$^{-3}$ and typical sizes $\Delta_{cl}= 0.001\times
 x_0 = 0.3x_0 /\Gamma_0 = 10^{13}$ cm, so that their angular extent is
 smaller than the Doppler angle. The blast wave is assumed to form a
 thin shell with $\eta = 1/\Gamma_0$. In an idealized circumburst
 medium where cloud properties were roughly the same throughout this
 region, then for every bright pulse or two, there would be another
 $\sim 7$ dimmer by a factor of 10 -- 20, and another 20 pulses dimmer
 than the brightest pulse by a factor $\approx 100$.  Pulse widths of
 a second or so are formed by the brightest pulses for such clouds, in
 keeping with the pulse paradigm \citep{nor96}.  At late times and
 from larger angles a lower fluence emission plateau would be formed
 if the jet and clouds are assumed to maintain roughly uniform
 properties at these angles.

Because of the unknown distribution of cloud densities, sizes and
locations, as well as effects of GRB jet structure, simulating GRB
light curves within an external shock scenario is very
model-dependent.  But simulations of GRB light curves with small
clouds randomly distributed in a shell show many of the features of
spiky GRB light curves \citep[][and below, Section 4.5]{dm99,dm04}.

The calculations shown here mean that given the thin-shell
approximation $\eta \cong 1/\Gamma_0$, essentially all BATSE GRBs
without redshift information, and many BATSE/Beppo-SAX/HETE-2 GRBs
with redshift information can be modeled within an external shock
scenario. The standard jet properties are $E_0 = 10^{53}$ ergs and $z
= 1$, in accord with pre-Swift observations of GRBs with redshift
information \citep{fb05}.  These pulses have SEDs in agreement with
BATSE observations \citep[e.g.][]{pre00}, bearing in mind different
ways to remedy the ``line-of-death" problem
\citep{pre98} in a nonthermal synchrotron-shock model through the
jitter mechanism \citep{med00}, radiation reprocessing \citep{db00},
etc.\ \citep[see][for a discussion of GRB spectral break models]{zm02}.

\begin{figure}[t]
\epsscale{1.0}
\plotone{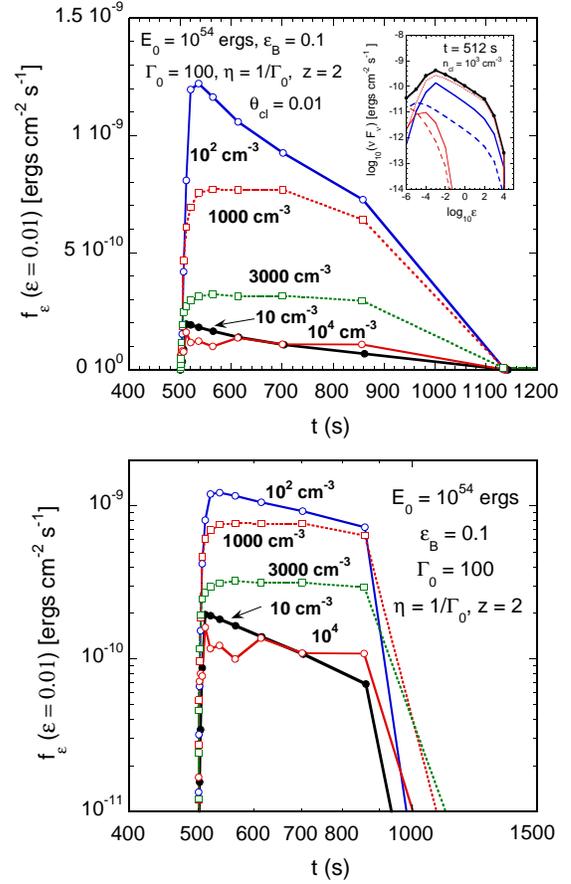}
\caption{Light curves calculated at $\e = 0.01$ ($h\nu \approx 5$ keV)
 from a blast-wave shell in the thin-shell approximation
$\eta = 1/\G_0$ for a blast-wave Lorentz factor $\Gamma_0 = 100$
 impacting an on-axis stationary cloud with opening angle $\theta_{cl}
 = 0.01$.  Top figure: linear flare profiles as a function of
 $n_{cl}$. Inset: SED measured at observer time of 512 s with $n_{cl} =
 10^3$ cm$^{-3}$. Bottom figure: logarithmic flare profiles as
 a function of $n_{cl}$.  }
\label{f9}
\end{figure}

Many X-ray flares observed with Swift can also be modeled within the
thin-shell approximation, as illustrated by the calculation shown in
Fig.\ 9. Here the standard X-ray flare parameters from Table 1 are
used. The principal difference between these parameters, which apply
to Swift observations, and those for the BATSE GRB pulse observations
is that the source redshift is $z = 2$ and the apparent isotropic
energy release is $10^{54}$ ergs, as implied by observations of GRBs
at such redshifts \citep{ghi07}.  The flares were detected by the XRT
on Swift, so that the mean telescope photon energy is $\epsilon
\lesssim 0.01$.  As can be seen, flares at flux levels reaching
$10^{-9}$ ergs cm$^{-2}$ s$^{-1}$ can be made by external shocks.
Comparison with observations of GRBs with measured redshifts in the
Swift catalog \citep{obr06} shows that most flares are well below this
level, except for a flare from GRB 050820A at $z = 2.612$ that reached
a $\nu F_\nu$ flux between 0.3 and 10 keV of $\approx 10^{-8}$ ergs
cm$^{-2}$ between 200 and 300 s after the GRB.

The SED for the model GRB X-ray flare peaking hundreds of seconds
after the start of the GRB is shown by the inset in Fig.\ 9.  In these
flares the reverse-shock synchrotron component can make an enhanced
optical/IR emission component due to the concavity produced by the
various RS components.

GRBs which lack redshift information, such as GRB 050502B with its
enormous X-ray flare exceeding $10^{-9}$ ergs cm$^{-2}$
\citep{fal06a}, pose no difficulty to the external shock
model.\footnote{Because the $\nu F_\nu$ peak flux was above the Swift
BAT energy window, the fraction of total GRB fluence in the X-ray
flare is unknown.} If the frozen pulse assumption is allowed, then
there is no difficulty in explaining $\gamma$-ray pulses or X-ray
flares in GRBs, provided that the surrounding medium of a GRB is
clumpy on well-defined density and size scales.

\begin{figure}[t]
\epsscale{1.0}
\plotone{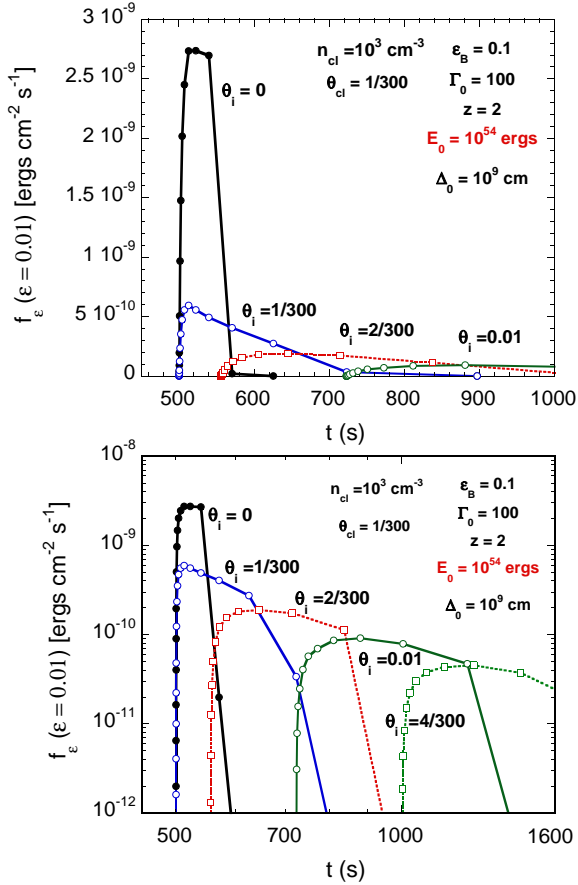}
\caption{Light curves calculated at $\e =
0.01$ ($h\nu \approx 5$ keV) from a blast-wave shell/stationary
cloud interaction in the frozen-pulse approximation $\eta \rightarrow
0$ for a blast-wave Lorentz factor $\Gamma_0 = 300$ impacting an
on-axis cloud with opening angle $\theta_{cl} = 1/300$ and $\Delta_0 =
10^9$ cm.  Top figure: linear flare profiles as a function of
incidence angle $\theta_i$ for $n_{cl} = 10^3$ cm$^{-3}$. 
Bottom figure: logarithmic flare profiles.  }
\label{f10}
\end{figure}

A straightforward prediction of the external shock model
as presented here is that the spectra of X-ray flares tend
to become softer with time, and will be systematically softer
than the pulses in the prompt phase. This is because the single 
expanding blast wave becomes less dense with radius, so that
interactions with surrounding cloud material 
tend to have weaker forward shocks. Thus there will be a 
systematic reduction of the $\nu F_\nu$ peak energies 
and a softening of the X-ray flare spectra. 
Softening of spectra are observed in flares from individual
GRBs like GRB 050502B \citep{fal06a} or GRB 
050822 \citep{god07}. This cannot be considered a definitive
test of the external shock model, however, because such behavior
can also be accounted for by a 
colliding shell/internal shock model if the relativistic
ejecta tend to have lower $\Gamma$ factors or less energy with time.
It would be surprising and contrary to expectations of this model
if X-ray flares tend to be harder than the prompt emission.

\section{Physics of Pulse and Flare Formation in the External Shock Model}

We find from this analysis that short pulses or flares can be made if
the GRB blast wave undergoes no transverse spreading and interacts
with small dense ($\sim 10^3$ cm$^{-3}$) clouds within $\approx
10^{16}$ -- $10^{17}$ cm from the GRB explosion.  The formation of
strong forward shocks with $\Gamma\cong \Gamma_0$ is necessary to make
bright pulses, and was the underlying assumption of the GRB light
curve simulations in \citet{dm99,dm04}, where GRB light curves similar
to actual GRB light curves were simulated with $\sim 10$\% efficiency.
The implications and plausibility of the assumption of a thin blast
wave shell are now considered.

\subsection{Temporal Behavior}

The characteristic timescales shown in the preceding figures can be
understood from simple considerations \citep{fmn96,sp97,dm99,
iok05}. Eq.\ (\ref{tz}) gives the time
\begin{equation}
t \;=\;{(1+z) x\over \beta_0 c} \;(1-\beta_0 \mu_i)\;,
\label{ti1z}
\end{equation}
since the start of the GRB explosion when emission is first received
from a cloud at location $x$ and inclination angle $\theta_i =
\arccos\mu_i$ that is impacted by a blast wave moving with Lorentz
factor $\Gamma_0$.  The angular timescale giving the temporal duration
over which emission, emitted at radius $x$ in the angular range
$\theta_2 = \theta_i+ \theta_{cl} > \theta_1 = \max(0,
\theta_i - \theta_{cl})$,
is received is simply
$${\Delta t^{ang}\over 1+z} \;=
\;{x\over \beta_0 c} (\mu_1 - \mu_2) \;\;
\stackrel{\theta_1,\,\theta_2\ll 1,~\Gamma_0\gg 1}{\rightarrow}\;\;
$$
\begin{equation}
\;\cong \; {x\over 2 c} \;(\theta_2^2-\theta_1^2)\;\cong ({x\over 2 c}) 
\;4\theta_i\theta_{cl}
\label{ti1z1}
\end{equation}
when $\theta_i > \theta_{cl}$, and 
\begin{equation}
{\Delta t^{ang}\over 1+z} \;
\;\cong \; {x\over 2 c} \;(\theta_i^2-\theta_{cl}^2)\;
\label{ti1z2}
\end{equation}
when $\theta_i <\theta_{cl}$.

The radial timescale for emission from two points at different
distances $x_1$ and $x_2$ but at the same inclination angle $\theta_i$
is given from eq.\ (\ref{tz0}) by
\begin{equation}
{\Delta t^{rad}\over 1+z} = {\Delta_{cl}\over \beta_0 c}\;
(1-\beta_0\mu_i ) \cong t_{min}\; 
\left( \Delta_{cl}\over x\right) \;(1+\theta_i^2\Gamma_0^2) \;,
\label{Deltatrad}
\end{equation}
where $t_{min}$ is defined by eq.\ (\ref{tmin}),  
and the inclination and cloud angles are in units of the Doppler angle
$\theta_0 = 1/\Gamma_0$ by the relations $$ {\rm v}_i= \Gamma_0
\theta_i\;\; {\rm and}\;\; {\rm v_{cl}} = \theta_{cl}\Gamma_0\;,$$
respectively.  The characteristic duration of a shell/cloud collision
is therefore 
$$\Delta t \cong t_{min}\;\max[{2{\rm v}_{cl}\over
\Gamma}(1+{\rm v}_i^2 ),$$
\begin{equation}
4{\rm v}_i{\rm v}_{cl}H({\rm v}_i - {\rm v}_{cl}) + 
({\rm v}_i + {\rm v}_{cl})^2 H({\rm v}_{cl}-{\rm v}_i)]\;,
\label{deltattotal}
\end{equation}
where $H(u) = 1$ if $u\geq 1$ and $H(u) = 0$ otherwise.
Note that the angular timescale is independent of cloud location $x$,
using ${\rm v}_{cl} = \Gamma_0\theta_{cl} = (\Delta_{cl}/x)\Gamma_0$
in eq.\ (\ref{deltattotal}), but depends on cloud size $\Delta_{cl}$
and inclination angle $\theta_i$. The radial timescale is also
independent of $x$ when written in terms of cloud size.

The STV condition for the external shock model is ${\rm v}_{cl} \ll
1$. Thus we see that the kinematic duration is determined by the
angular timescale everywhere except when ${\rm v}_i \leq 1/2\Gamma_0$,
that is, for nearly on-axis clouds with $\theta_i \leq 1/2\Gamma_0^2$.
The kinematic duration of the pulse $\Delta t\propto 1/\theta_i$ for
$\theta_i \geq 1/2\Gamma_0^2$. The pulse shortening for nearly on-axis
events, combined with beaming factors and spectral effects, is shown
from analyses and Monte Carlo simulations capable of explaining the
range of observed GRB BATSE light curves \citep{dm99,dm04}.

\subsection{Optimal Cloud Density for Bright Pulses}

The optimal density giving the brightest measured flux can be derived
from three requirements:
\begin{enumerate}
\item A strong FS which, from eq.\ (\ref{n(x)/n}), implies a maximum
cloud density given by
\begin{equation}
n_{cl}  \lesssim {E_0 \over 16\pi x^2 \Gamma_0^4 m_p c^2 \Delta(x)}\;.
\label{nclmax}
\end{equation}
\item Significant blast-wave deceleration to provide efficient energy
extraction, which requires clouds with thick columns
\citep{dm99}, that is, with densities
\begin{equation}
n_{cl}  \gtrsim {E_0 \over 4\pi x^2 \Gamma_0^2 m_p c^2 \Delta_{cl}}\;.
\label{nclmin}
\end{equation}
\item The requirement of a strong FS and a thick column therefore
translates into the requirement that
\begin{equation}
\Delta_{cl} \gtrsim 4\G_0^2 \Delta(x)\;
\label{deltacl}
\end{equation}
in order to produce STV. The STV condition is
\begin{equation}
 \theta_{cl} = {\Delta_{cl}\over x} = 
{{\rm v}_{cl}\over \Gamma_0} \ll {1\over \Gamma_0} \;.
\label{Deltacl}
\end{equation} 
\end{enumerate}
 
Using eq.\ (\ref{Delta(x)}) for the shell width, eq.\
(\ref{deltacl}) becomes 
$$
4\eta\Gamma_0 < {4\G_0^3 \Delta_0\over  x } + 4\eta \G_0  \ll 1\;,$$ 
so that the requirement on the width 
of the radiating shell material in
the external shock model for rapid variability is simply that 
\begin{equation}
\eta \lesssim {1\over 4\Gamma_0} \;\;{\rm when}\;\; 
x\gtrsim 4\Gamma_0^3 \Delta_0 \cong 10^{15} \Gamma_{300}^3 \Delta_7\;
{\rm cm}\;.
\label{etamax}
\end{equation}
It is therefore not possible to have STV if $\eta \cong 1$, because
then the reverse shock is relativistic, the forward shock Lorentz
factor is low and Doppler boosting is lost.  But it becomes possible
in the thin-shell approximation $\eta \sim 1/\G_0$. As shown by the
light curves and SED calculations, the phenomenology of GRBs can be
explained in the frozen pulse ($\eta = 0$) and also, possibly, the
thin-shell approximation.

The optimal cloud density for bright variable pulses in the external
shock model, from eq.\ (\ref{nclmax}) for a GRB blast wave with
apparent isotropic energy release $E_0 = 10^{54} E_{54}$ ergs and
shell width $\Delta(x) = 10^7\Delta_7$ cm that encounters a cloud at
$10^{16}x_{16}$ cm from the center of the GRB explosion, is
\begin{equation}
n_{cl}({\rm cm}^{-3})  \approx 1.6\times 10^6 \;{E_{54} \over 
 x_{16}^2 \Gamma_{300}^2 \Delta_{7}}\;.
\label{ncloptimal}
\end{equation}
For instance, the calculations shown in Fig.\ 5 with $\eta = 1/\Gamma_0$ 
have $\Delta_7(x=10^{16}{\rm~cm}) = 38$.

\subsection{Width of Radiating Shell}

Besides the requirement for STV that the unshocked fluid shell be thin
or frozen, the width of the shocked fluid shell cannot be too great,
for otherwise rapid variability would be washed-out by contributions
from different parts of the shell.

First consider a blast wave sweeping up matter from a uniform
surrounding circumburst medium with density $n_0$(cm$^{-3}$). The
swept-up material, compressed by the factor $4\Gamma$ at the external
shock, has thickness
\begin{equation}
\Delta_{sh}^\prime(x) = {x\over 12\Gamma}\;\;{\rm and}\;\; 
\Delta_{sh}(x) = {x\over 12\Gamma^2}\;
\label{deltaprimex}
\end{equation}
in the proper and explosion frames, respectively.
Because the angular extent of the shocked fluid layer in 
the blast wave covers the full Doppler cone, the variability 
is already limited to a timescale $\gtrsim 0.25 t_{min}$ by 
light travel time limitations. Under these circumstances, 
rapid variability is not expected, even when considering
density jumps \citep{ng06}.

Now consider the transverse width of the shocked fluid shell colliding
with a cloud with Dopper size $v_{cl}\ll 1$ that subtends the solid
angle $\Delta \Omega_{cl}$ as seen from the center of the
explosion. By equating the mass $m_p \Delta \Omega_{cl} x^2 n
(x) \Delta_{sh}^\prime (x)$ with the cloud mass $m_p \Delta
\Omega_{cl} x^2 n_{cl} \Delta_{cl}$ we have, using the relativistic
density jump condition $n(x) = 4\Gamma n_0$,
\begin{equation}
\Delta_{sh}^\prime(x) = {\Delta_{cl}\over 4\Gamma} = 
{{\rm v}_{cl}x\over 4\Gamma^2}\;\;
{\rm and}
\;\; 
\Delta_{sh}(x) = {{\rm v}_{cl}x\over 4\Gamma^3}\;,
\label{deltaprimex1}
\end{equation}
i.e., a shocked-fluid layer thinner than required
to produce STV.

\subsection{Thin Shell Assumption}

From the preceding discussion, one can see that the viability of an
external shock model for the $\gamma$-ray pulses and X-ray flares
depends on whether the GRB blast-wave width spreads in the coasting
phase according to eq.\ (\ref{Delta(x)}), with $\eta \lesssim
1/\Gamma_0$.  In the gas-dynamical study of \citet{mlr93},
inhomogeneities in the GRB fireball produce a spread in particle
velocities of order $|v - c|/c \sim \Gamma_0^{-2}$, so that $\Delta(x)
\sim x/\Gamma_0^2$ when $x \gtrsim \Gamma_0^2 \Delta_0$ and $\eta\sim
1$. This expression was also obtained in the hydrodynamical analysis
by \citet{psn93}.

Several points can be made about these results. First, gas-dynamical
or hydrodynamic analyses omit the fluid nature of the explosion plasma
whose behavior depends crucially on the magnetic field and MHD
properties of the fluid.  Second, the spread in $\Delta$ considered
for a spherical fireball is averaged over all directions.  As the
fireball expands and becomes transparent, the variation in fluid
motions or gas particle directions over a small solid angle $\sim
1/\Gamma_0^2$ of the full sky becomes substantially less.  Small
values of $\eta$ could be realized if there is little dispersion in
the baryon-loading of the fluid shell within the small range of solid
angles subtended by the jet from which the observed emission is
radiated.  Third, the particles within the cold unshocked blast-wave
shell will expand and adiabatically cool so that the fluid will spread
with thermal speed $v_{th}= \beta_{th} c$. The comoving width of the
blast wave is $\Gamma_0 \Delta_0 + \beta_{th} c \Delta \tp \approx
\Gamma_0 \Delta_0 + \beta_{th} x /\Gamma_0$, so that the 
spreading radius $x_{spr} \cong \Gamma_0^2 \Delta_0/\beta_{th}$.
Adiabatic expansion of nonrelativistic particles 
can produce a very cold shell with
$\beta_0 \lesssim 10^{-3}$, leading to very small shell 
widths. 

The requirement on the thinness of $\Delta(x)$ does not apply to the
adiabatic self-similar phase, where the width of the shocked fluid
shell is $\Delta_{sh} \sim x/12\Gamma_0^2$, as implied by the
relativistic shock jump equations \citep{bm76}; see eq.\
(\ref{deltaprimex}). Consequently, no extreme variability is expected
when a blast wave, already in the adiabatic self-similar regime,
encounters a density jump, as demonstrated by \citet{ng06} (in
agreement with the results of unpublished numerical simulations using
the code described by \citet{cd99}). But even in this case, however,
$\Delta \ll x/\Gamma_0^2$ if the blast wave is highly radiative
\citep{cps98}, which could result from leptonic processes when $\e_e,
\e_B \gtrsim 0.1$, or hadronic processes when $\e_e \sim 0.1$ and
$\e_B \gtrsim 0.1$ and the surrounding circumburst medium is dense,
with $n_0 \gtrsim 10^3$ cm$^{-3}$ \citep{der06}.

This question deserves considerably more study, and will require at
least an MHD model employing self-consistent particle distributions
and magnetic field geometries in the erupting plasma, generation and
amplification of the magnetic field in the shocked fluid layer
(including particle acceleration and diffusion into the cold fluid
shell), and magnetic field and particle energy evolution from shell
expansion.

\subsection{Density Contrast}

A further requirement on this model is that the dense clouds are found in 
a medium sufficiently tenuous that the blast wave has not undergone significant
deceleration by sweeping
up this material. This constraint can be expressed by the condition that 
the deceleration radius $x_d = (3E_0/4\pi m_pc^2 n_0 \Gamma_0^2)^{1/3}$ \citep{rm92,mr93}
be much greater than the radius defining the condition for a thick column, eq.\ 
(\ref{nclmin}). This translates to the requirement that the density contrast between
the surrounding medium density $n_0$ and the cloud density $n_{cl}$ is given by
$${n_0\over n_{cl}} \ll \Gamma_0\;\sqrt{{36\pi n_{cl} \Delta_{cl}^3\over E_0/m_pc^2}}
$$
\begin{equation}
\sim 10^{-10}\;\Gamma_{300} \; \sqrt{(n_{cl}/10^3{\rm~cm}^{3})(\Delta_{cl}/10^9{\rm~cm})^3\over E_{54}}\;
\label{n0ncl}
\end{equation}
This is a severe constraint that can only be realized in an extreme environment such as the 
one described in more detail below.

\subsection{Simulations of GRB Light Curves}

\begin{figure}[t]
\epsscale{1.1}
\plotone{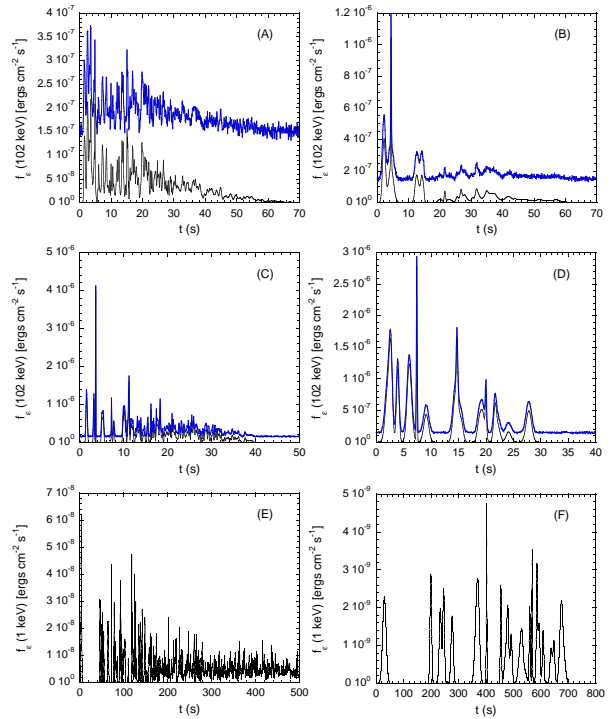}
\caption{Simulated GRB lightcurves from the interaction of a thin
blast wave with clouds with thick columns, using parameters from Table
2. The top four figures show $\approx 100$ keV 
light curves. The lower curve is the GRB emission and
the upper curve is this emission added to Gaussian noise typical of
the BATSE detectors. The bottom two figures show light curves at 5 keV characteristic
of X-ray emission detected with the XRT on Swift.}
\label{f11}
\end{figure}

Fig.\ 11 shows Monte Carlo simulations, described in more detail in
\citet{dm04}, of GRB light curves under the assumption that the cold
blast wave shell remains thin and the interaction forms a strong
forward shock. The parameters of the simulation are shown in Table 2,
and here we use fixed cloud sizes with radius $r_1$ that are
``uniformly randomly" distributed in the volume between $R_1$ and
$R_2$. In all cases, the surface filling factor $\xi = 10$\%, and a
fraction $\zeta = 10$\% of the blast wave energy intercepted by the
cloud is transformed into radiation received by the detector at energy
$m_ec^2 \e$.  The angular extent of the jet is $k_\theta/\Gamma_0$,
and an occulting factor is included so that the portion of the blast
wave that intercepts a cloud at a smaller radius no longer radiates
when interacting with clouds at larger radii.

The top four figures represent the GRB prompt phase detected near 100
keV using parameters appropriate to BATSE, including noise at a level
typical of the BATSE detectors.  The bottom two figures make use of
parameters that would apply to Swift.  The variety of simulated light
curves is endless, depending on the seed used in the random number
generator and the choices of cloud and blast wave
parameters. Precursor events can  be made if some material is
found very close to the GRB \citep[][compare figure E]{piro05}. 

\subsection{Criticisms of the External Shock Model}

We describe the general criticisms that the external shock model for
prompt GRB emissions has endured.  \citet{fmn96} demonstrated that an
expanding blast wave would, if briefly illuminated over a large
portion of its surface covering the Doppler cone, make progressively
longer pulses in accordance with the relation $t\cong (1+z) x/\Gamma^2
c$. The tendency for pulse lengthening can be avoided if a small
fraction of the blast wave surface area within the Doppler cone were
illuminated at any one times, but this was thought to require an
unreasonable number of illuminating regions (``clouds").

One approach is to estimate a maximum ``surface filling factor" of
radiating sites consistent with measured variability, using kinematic
arguments \citep{fen99}.  \citet{sp97} argued that highly variable
light curves cannot be constructed from blast-wave interactions with
surrounding density inhomogeneities by superposing pulses made near
the outer region of the Doppler cone to show how they overlap due to
the large angular timescales; hence the sources of GRBs must require
an active central engine to make discrete pulses. \citet{dm04} showed
that this argument, made also more recently by \citet{pir05} and
\citet{iok05}, mistakenly uses the angular time scale near the outer
edge of the Doppler cone, and misses a number of important points: One
is that the beaming factors, not considered in these papers, heavily
weight the total fluence for near on-axis clouds, and even more so the
pulse peak flux because of the smaller kinematic variability timescale
for nearly on-axis clouds; another is the different (observer) times
for the clouds nearly on-axis as compared with those off-axis, so that
flux ratios from different parts of the shell are time-dependent; a
third is the assumption that clouds are distributed with azimuthal
symmetry within the Doppler cone; a fourth is that the beaming factor
of the relativistic jet that forms $\gamma$-ray pulses in the prompt
phase may be smaller than $\sim \Gamma_0^{-1}$ and the typical angles
inferred from optical afterglow breaks, which will reduce the
importance of off-axis events. The crucial role of the narrow cold
blast-wave fluid shell (not the shocked fluid shell considered by
\citet{ng06}) to ensure a strong forward shock has also been overlooked in
past studies.

\citet{rf00} claim that the lack of pulse-width evolution in BATSE 
GRB light curves favors internal shock models; \citet{kbb07}, in the
absence of any quantitative consideration of external shocks, argue
that XRT pulse spreading in Swift GRBs favors internal shock models.
From our simulations, we see that this effect is subtle at best in the
unlikely condition when the clouds are ``uniformly randomly" situated.
Realistic cloud clustering properties or a narrow $\gamma$-ray jet
could obscure this effect completely. \citet{lp07} derive the
time-scale ratio $\Delta t_v/\hat t \sim 0.25$ for a spherical shell
and suggest that the observational data indicate a clustering in the
value $\Delta t_v/\hat t \sim 0.25$ that rules out the external shock
model, but nowhere consider observational biases for the detection of
flare profiles with larger or smaller time-scale ratios, or whether
this time-scale ratio could reflect cloud properties.

As shown in the simulations, STV in the external shock model with
moderate ($\sim 10$\%) efficiency is possible if the blast wave
interacts with clouds with thick columns and sizes $\Delta_{cl} \ll
x/\Gamma_0$, because then the condition of ``local spherical symmetry"
\citep{fmn96} is broken. The shortest pulses carrying a significant
fraction of the total fluence are made by interactions of the blast
wave with small clouds at angles $\theta \ll 1/\Gamma_0$ to the
observer line-of-sight \citep{dm04}. Efficiency concerns are reduced
if the fraction of energy dissipated as X-rays and $\gamma$ rays is a
small fraction of the total
\citep{der06}.

\section{Discussion }

We have critically examined whether external shock processes can make
short timescale X-ray and $\gamma$ ray variability in the $\gamma$-ray
luminous phase and the early afterglow phase, extending to $\gtrsim
10^4$ s after a GRB trigger. The result is simple: external shocks can
only make the erratic variability observed in GRB light curves if the
cold fluid shell ejected by the impulsive burst event does not spread
significantly from its initial characteristic width $\Delta t_{exp}/c$
at the time of the GRB explosion, where $\Delta t_{exp}$, assumed to
be $ \ll 1$ s, is the time of the energy release in the explosion
frame. The question of whether lack of shell spreading is a valid
assumption, at least within the Doppler cone primarily sampled by the
observer, is an open question but plausibly valid.  A crucial point is
that an external shock model for the rapid variability is feasible
because the GRB blast wave is still in its coasting phase, and has not
yet reached the adiabatic self-similar phase.

Let us follow the implications if this assumption were allowed. 

One way to interpret GRB light curve data is to suppose that there is
a single impulsive burst of energy, and all subsequent phenomenology
is a consequence of the collimated ultra-relativistic blast wave
interacting with its surroundings \citep{mr93,dm99,dm04}.  External
shocks are the basis for the success that the relativistic blast-wave
physics model has had in interpreting GRB afterglows
\citep{mr97,zm04}; we also suppose that its success can be translated
to the prompt phase. The properties of the surrounding medium, as
implied by extensive evidence on GRB host galaxy studies, is
determined by the type of GRB progenitor, which is a massive ($\gtrsim
10$ -- 20 M$_\odot$) star lacking a H envelope whose core collapses to
form a Type 1b/c SN. The SNR and the progenitor stellar wind,
including effects of a possible binary companion \citep{fry07}, form a
complicated surroundings that imprints its structure on the GRB light
curves \citep{dm99,wl00}.

The brief impulsive energy releases in GRB explosions, which for
neutrino-mediated processes are most efficient for $\simeq$ ms
timescales, are compatible with brief collapse events or phase
transitions from stellar cores to compact objects, for example, from
neutron star to black hole or neutron star to a quark star
\citep[e.g.,][]{dlp05,ber03}.  This collapse event would follow by
minutes, hours, days or more a Type 1b/c supernova that forms a
supramassive, rotationally supported neutron star/magnetar.  This
picture is a two-step collapse model, like the supranova model of
\citet{vs98,vs99}, though with a short delay.

Such a picture, orthogonal in all respects to the collapsar model
except for the primary role of high mass stars as long-duration GRB
progenitors, has a great deal of explanatory and predictive
capability.  Here we point out some of the attractive features of such
a picture:
\begin{enumerate}
\item $^{56}$Ni. The direct collapse of the evolved core of a massive 
star to a black hole in the collapsar model, avoiding core bounce,
requires new routes for the production of $^{56}$Ni. This makes the
similarity between the peak optical rest frame $M_V$ magnitudes (taken
as a proxy for $^{56}$Ni production) measured in SNe associated with
GRBs and Type 1b/c SNe not associated with GRBs \citep{sod05} a
remarkable coincidence. Two-step collapse processes avoid this
coincidence by forming $^{56}$Ni through neutron-star core bounce.

 \item Energy reservoir. The discovery \citep{fra01,pk01,fb05} of a
 standard energy reservoir, or at least a clustering of
 beaming-corrected absolute energies towards a value $\cong 10^{51}$
 ergs, modulo an uncertain efficiency to convert explosion energy into
 $\gamma$ rays, fixes the energy budget of the GRB explosion.  The
 collapse of a rotationally-supported neutron star to a black hole, or
 the transition from a degenerate neutron fluid to quark state, could
 liberate kinetic energies approaching a constant maximum energy that
 may not defy ability or require excessive assumptions to
 calculate. If the explosion involves a neutrino or Poynting-flux
 mediated interaction, values of collimated electromagnetic energy of
 $\approx 10^{51}$ -- $10^{52}$ ergs could be released in a second
 collapse to a black hole.

\item Complex circumburst medium.
A magnetar phase \citep{uso92} follows, even if only by minutes, the
first collapse event, and drives a strong pulsar wind to disrupt,
thread, and render extremely complex the surrounding
medium, in particular, the supernova remnant shell produced in the 
first collapse event.
The work of \cite{kg02} shows the profound impact on the circumburst medium that
this phase can have for long (month--yr) delays.

\item Clean environment. The Usov phase cleans the environment except for 
dense clumps formed at distances where the strong radiation pressure
can be withstood---and these are the clumps that form the clouds in
the external shock model. The MHD winds effectively drive out all surrounding
material except for very dense clouds, possibly allowing
for the existence of density contrasts of magnitude defined by eq.\ (\ref{n0ncl}. 
The clouds themselves would be expanding, but on a much longer timescale than 
the period of activity of the system. Close to the explosion trigger, the
environment of a rotationally-supported neutron star is baryon-clean
\citep{vs98}. 

\item Explanation for chromatic X-ray and optical afterglow beaming breaks.
\citet{pan06} compare Swift BAT X-ray light curves with optical light 
curves taken from a variety of telescopes for 6 GRBs, finding several
instances where the X-ray light curves appear to decay chromatically
with respect to the optical light curves.  Possible explanations for
these observations are evolving $\e_e$ and $\e_B$ parameters and the
addition of the (synchrotron self-)Compton for a beamed outflow
\citep[illustrated numerically by][]{dcm00}, which plausibly explains
the light curves of GRB 050922C and the lack of X-ray breaks observed
with Swift \citep{sat07}.  The optical light curves of GRBs 050607,
050802, and 050713A are too poorly sampled to draw definitive
conclusions, and the optical and X-ray light curves in GRB 050319
appear to follow each other except for a final high optical datum.
This leaves only GRB 050401 to explain. More and better data is
clearly needed.

\item Explanation for the decay and plateau phases observed with Swift.
Within the context of the external shock model for a GRB taking place
within a uniform surrounding environment, I recently showed that an
intense photohadronic phase, effectively depleting internal GRB blast
wave energy, could cause the X-ray light curves to exhibit rapid flux
declines \citep{der06}. After the discharge, the blast wave regains
its nonradiative character to form the plateaus observed with Swift.
Because GRBs showing a steep decay phase have significantly decelerated, the 
blast wave should then have entered the adiabatic self-similar phase
when only broad X-ray flares can result from large density contrasts
\citep{ng06}. It is interesting to note that GRBs showing the most
extreme decay phases, e.g., GRBs 050315, 050416A, 050713B, 050814, 
and 050915B, display smooth subsequent light curves \citep{obr06,chi07}, 
though GRB 050916 provides
a counterexample and a challenge to this interpretation.

\item High energy radiation.
Observations of GRBs with the GLAST GBM and LAT will monitor the
Compton components in the spectrum of a GRB.  Definite correlations
between the leptonic synchrotron and SSC components are expected,
which behave in stark contrast to photohadronic $\gamma$-ray
components that vary independently of the lower-energy lepton
synchrotron component. GLAST will search for photohadronic emission
components and, in conjunction with detectors such as IceCube, test
multiwavelength and multi-channel predictions \citep{drl07}.

\item Predictions. With respect to the short-delay supranova model, 
the confirming prediction is to glimpse the GRB progenitor in its
magnetar phase prior to the second-step collapse to form the GRB.
Analysis of radio, optical, X-ray, or $\gamma$-ray emissions to obtain
$P(t)$ and $\dot P(t)$ in the brief interval between the two collapse
events is now only possible with broad field-of-view detectors with
poor sensitivities.  Unfortunately, optical supernova discovery
generally happens within a few weeks after the supernova event, by
which time a second collapse would have happened.  So for the
short-delay supranova intervals consistent with the offset between
nearby GRBs and their supernova emissions \citep{zkh04}, this
prediction is not promising except to test the standard supranova
model \citep{vs98}.
 
A definitive test of the collapsar model is to find a supernova taking
place well in advance of the GRB, so that the GRB has no detectable
supernova emissions.  Development of telescopes with wide
field-of-view optical survey capabilities, like Pan-STARRS or SNAP,
holds promise to rule out the collapsar model for specific GRBs (C.\
Dermer \& A.\ Drago, unpublished, 2006), and could also provide a
database to search for magnetar activity in advance of a GRB.

\end{enumerate}

\section{Summary and Conclusions}

A detailed analysis of the interaction between a relativistic blast
wave and a stationary cloud was performed in the limit that that the
cloud width $\Delta_{cl} \ll x$, so that the shell density remains
effectively constant during the interaction.  Synchrotron light curves
and spectra from such interactions were calculated for a range of
cloud densities, sizes, and locations, for blast-wave coasting Lorentz
factors $\Gamma_0 = 100$ and 300, and for different blast-wave widths.

In order to produce short timescale variability in the external shock
model, the shell width parameter $\eta$ must be $\lesssim 1/\Gamma_0$,
as demonstrated both analytically and numerically. If this model
assumption is valid, then the external shock model can explain the
generic spectral shape of GRB pulses, rapid variability in GRB light
curves observed with BATSE, and the delayed X-ray flares observed with
Swift. The tendency of the GRB pulses or X-ray flares to diminish in
intensity with time is a consequence of the expansion of the blast
wave. The durations of GRB pulses reflect the dimensions of the
surrounding thick-columned clouds.  The typical distances of the
clouds that produce $\gamma$-ray pulses are $\approx 10^{15}$ --
$10^{17}$ cm from the GRB explosion center, with cloud sizes $\simeq
10^{12}$ -- $10^{13}$ cm and densities $\sim 10^3$ cm$^{-3}$.  The
clouds that produce X-ray flares observed with Swift are typically
found $\gtrsim 10^{17}$ cm from the center of the GRB explosion, and
have sizes $\lesssim 10^{15}$ cm and densities $\sim 10^3$ cm$^{-3}$.

This analysis also provides a basis for making accurate calculations of
light curves and spectra formed by both forward and reverse shocks
in collisions between relativistic shells.

\acknowledgements

I would like to thank A.\ Atoyan, D.\ Band, M.\ B\"ottcher, C.\ Fryer,
M.\ Gonz\'alez, J.\ Granot, E.\ Nakar,  E.\ Ramirez-Ruiz, and B.\ Zhang for
discussions. I acknowledge very important points made by the referee
of this paper. I would also like to thank Kurt Mitman for the use
of the numerical simulation code that we developed to study this
problem, and to thank Alessandro Drago for his visit.  This work was supported
by the Office of Naval Research. I gratefully acknowledge GLAST
Interdisciplinary Science Investigator grant DPR S-13756G and Swift
Guest Investigator grants, without which this research would not be
possible.


\begin{deluxetable}{lll}
\tablecolumns{2}
\tablewidth{0pt}
\tablecaption{Standard Blast Wave and Cloud Parameters\tablenotemark{a}}
\tablehead{& GRB pulse & X-ray Flare}
\startdata
Redshift $z$&$1.0$ & 2.0\\
$d_L(10^{28}\rm ~cm)$ & $2.02$ & 4.80\\
$E_0~{\rm ergs}$ & $10^{53}$ & $10^{54}$ \\
$\Gamma_0$ & $300$ & 100 \\
$\Delta_0$ & $10^{7} {\rm ~cm}$ & \\
$\eta $ & $\G_0^{-1},0
$ \\
$\e_e$ & $0.1$\\
$\e_B$ & $10^{-1},~10^{-4}$\\
$\e_2$ & $0.1$\\
$p$ & $2.5$ \\
$\tcl$ & $0.01$ \\
$\ti$ & $0.0$ \\
$x_0({\rm cm})$ & $10^{16}$ & $10^{17}$\\
$x_2({\rm cm})$ & $1.02\times 10^{16}$ & $1.02\times 10^{17}$ \\
$n_{cl}({\rm cm}^{-3})$ & $10^3$ & \\
\enddata
\tablenotetext{a}{Forward and reverse shock parameters are the same.}
\end{deluxetable}

\begin{deluxetable}{lcccccc}
\tablecolumns{6}
\tablewidth{0pt}
\tablecaption{Parameters for Simulated Light Curves\tablenotemark{a}}
\tablehead{& A & B & C & D & E & F }
\startdata
$E_0~{\rm (ergs)}$ & $10^{53}$ & & $10^{54}$  & & & \\
Redshift $z$&$1$ &  &  &  &  &  \\
$k_\theta $ & $1$ & &  0.5 &  &  &  \\
$\Gamma_0$ & $300$ &    &  &  &  & 100  \\
$\e$\tablenotemark{b} & $0.2$ &  &  &  & 0.002 &  \\
$\e_{pk}$\tablenotemark{b} & 0.4 &  &  &  &  &  \\
$R_1$(cm) & $10^{16}$ &  &  &  &  & $10^{17}$  \\
$R_2$(cm) & $10^{17}$ & &  &  & $10^{18}$ &  \\
$r_1$(cm) & $3\times 10^{12}$ & $ 10^{13}$ & $3\times 10^{12}$ & $ 10^{13}$ &  & $ 3\times10^{13}$ \\
$\xi$ & $0.1$ & &  &  &  &  \\
$\zeta$ & $0.1$ &  &  &  &  &  \\
\enddata
\tablenotetext{a}{Only the parameters changed from the previous column are listed.}
\tablenotetext{b}{Received photon energy and $\e_{pk}$ in units of $m_ec^2$.}
\end{deluxetable}

\end{document}